\documentclass[10pt,preprint]{aastex}

\newcommand{\degree}{\arcdeg}

\slugcomment{{\bf To appear in {\it The Astronomical Journal}, June 2004}}

\shorttitle{HIRES RBGS Atlas}
\shortauthors{Surace et al.}

\begin{document}

% titlepage

\title{An IRAS High Resolution Image Restoration (HIRES) Atlas of All 
Interacting Galaxies in the IRAS
Revised Bright Galaxy Sample}

\author{Jason A. Surace}
\affil{SIRTF Science Center, MS 220--6, California Institute of 
Technology, Jet Propulsion Laboratory, Pasadena, CA 91125}
\email{jason@ipac.caltech.edu} 

\author{D. B. Sanders}
\affil{University of Hawaii, Institute for Astronomy, 2680 Woodlawn Dr., 
Honolulu, HI, 96822}
\email{sanders@ifa.hawaii.edu}

\and

\author{Joseph M. Mazzarella }
\affil{Infrared Processing and Analysis Center, MS 100--22, California Institute of 
Technology, Jet Propulsion Laboratory, Pasadena, CA 91125}
\email{mazz@ipac.caltech.edu}

%abstract

\begin{abstract}

The importance of far-infrared observations in our understanding of
extreme activity in interacting and merging galaxies has been
illustrated by many studies.  Even though two decades have passed
since its launch, the most complete all-sky survey to date from which
far-infrared selected galaxy samples can be chosen is still that of the
Infrared Astronomical Satellite (IRAS).  However, the spatial
resolution of the IRAS all-sky survey is insufficient to resolve the
emission from individual galaxies in most interacting galaxy pairs,
and hence previous studies of their far-IR properties have had to
concentrate either on global system properties or on the properties of
very widely separated and weakly interacting pairs.  Using the HIRES
image reconstruction technique, it is possible to achieve a spatial
resolution ranging from 30\arcsec \ to 1.5\arcmin \ (depending on
wavelength and detector coverage), which is a fourfold improvement
over the normal resolution of IRAS. This is sufficient to resolve the
far-infrared emission from the individual galaxies in many interacting
systems detected by IRAS, which is very important for meaningful
comparisons with single, isolated galaxies.

We present high-resolution 12, 25, 60, and 100 $\mu$m images of 106 
interacting galaxy systems contained in the IRAS Revised Bright Galaxy 
Sample (RBGS, Sanders et al. 2003), a complete sample of all galaxies having a 
60$\mu$m flux density greater than 5.24 Jy.  These systems were selected 
to have at least two distinguishable galaxies separated by less than 
three average galactic diameters, and thus we have excluded very 
widely separated systems and very advanced mergers.  Additionally, 
some systems have been included which are more than three galactic 
diameters apart, yet have separations less than 4\arcmin, and which 
are thus likely to suffer from confusion in the RBGS.

The new complete survey has the same properties as the prototype
survey of \citet{hires}.  We find no increased tendency for
infrared-bright galaxies to be associated with other infrared bright
galaxies among the widely separated pairs studied here.  We find small
enhancements in far- infrared activity in multiple galaxy systems
relative to RBGS non-interacting galaxies with the same blue
luminosity distribution.  We also find no differences in infrared
activity (as measured by infrared color and luminosity) between late
and early-type spiral galaxies.

\end{abstract}


% mainbody

\section{Introduction}

In the last two decades it has become apparent that interactions 
between galaxies can play a significant role in their evolution.  From 
the early dynamical simulations of \citet{toomre} to more modern works 
by Barnes, Hernquist, and others (Barnes \& Hernquist 1992, and references 
therein) it has become apparent that interactions and mergers between 
galaxies can radically alter their morphology by inducing shells, 
bars, tails, and other tidal features.  Perhaps more importantly, 
cancellation of angular momentum during the merger process can lead to 
a radical redistribution of the gas content of the galaxies, with very 
rapid gas inflow into the galaxy cores.  This supply of fresh material 
could possibly fuel an active galactic nucleus, or provide the high 
gas densities needed to lead to a sudden burst of star formation.

There is considerable evidence that enhanced star formation is 
associated with interacting galaxies (Sulentic 1988, and references 
therein).  
The young OB stars that dominate the starburst radiate primarily in 
the optical and ultraviolet, but surrounding gas and dust reprocesses 
this radiation and thus strongly radiates at thermal wavelengths in 
the far-infrared.  Far-infrared luminosity is thus indicative of the 
magnitude of recent star formation activity \citep{telesco,lonsdale}.  
Additionally, due to the increased temperature of the heated dust, we 
expect the far-infrared colors to be a good diagnostic of enhanced 
star formation.  Therefore, many studies have therefore concentrated on the 
far-infrared properties of interacting galaxies.

Several studies have also discussed the incidence of multiple bright
galaxies being found within a given interacting galaxy system. 
\citet{haynes} found that for galaxy pairs separated by 2-10\arcmin,
approximately 10\% have multiple components brighter than 0.5 Jy at
60$\mu$m and 1 Jy at 100$\mu$m.  \citet{xu} also concluded that in the
majority of interacting systems, only one galaxy is infrared bright. 
These results agree with an earlier work by \citet{bob} which
concluded, based on near- IR colors, that most often only one galaxy
in a pair showed signs of unusual activity.  This is an interesting
result, because it suggests that specific properties of the
interacting galaxies may determine whether or not they become emitters
in the far-IR as well.  Testing this hypothesis requires resolution of
the individual galaxies in the far-IR, which is the goal of this IRAS
study.

The canonical figure used by many authors to delineate interacting 
versus non-interacting systems is a projected separation of three 
average galactic diameters, as presumably galaxies this close to one 
another are also close enough to exert a considerable gravitational 
effect \citep{dahari,byrd,hires}.  However, for most galaxies
detected by IRAS
this typically corresponds to an angular separation of a few 
arcminutes.  This is less than the resolution normally achieved by 
IRAS using the 1-d coadder ADDSCAN or the 2-d FRESCO imaging process.  
As a result, it has been impossible to study the far-infrared 
properties of the individual galaxies, and most studies have either 
made assumptions about the distribution of flux between galaxies 
within the interacting system \citep{bushouse} or have concentrated on 
widely separated pairs \citep{haynes,xu}.  Since previous studies of 
very widely separated galaxy pairs indicate that in a substantial 
fraction of interacting systems only one galaxy is unusually active in 
the far-IR \citep{xu}, it is necessary to resolve these 
galaxies in order to properly study those properties such as 
morphology which are unique to the individual galaxies.  Additionally, 
\citet{xu} found evidence that at smaller separations (and 
hence greater interaction strengths) there was a greater enhancement 
of far-IR activity.  Therefore it would be valuable if these studies 
could be extended to smaller separations where more observable changes 
are taking place.

Development of the Maximum Correlation Method algorithm (MCM, Aumann 
et al. 1990) for
use in IRAS image reconstruction significantly increased the resolution
of IRAS observations.  As implemented in the HIRES process, MCM is an
iterative image reconstruction technique that involves using the known
response functions of the IRAS detectors to scan simulated image
estimates which are then compared to the actual detector data.  In this
way, a high resolution image estimate is formed.  The result is
typically a five-fold increase in resolution, varying from roughly
30\arcsec$\times$45\arcsec \ at 12$\mu$m to 72\arcsec$\times$130\arcsec
\ at 100$\mu$m, with the actual achieved resolution being highly
dependent on the geometry of the detector coverage \citep{hires}.
Unfortunately, the HIRES process is extremely computer intensive.  
When developed, a single field
typically took a day or more to process.  As a result, the earlier work by
\citet{hires} was rather limited in scope, with only 23 systems being
resolvable.  On a modern computer, this computing time is
reduced to approximately 15 minutes, thus making feasible the
processing of a substantially larger sample
\footnote{HIRES processing is available from the Infrared Processing 
and Analysis Center Infrared Science Archive, Jet Propulsion Laboratory, California Institute 
of Technology {\it http://irsa.ipac.caltech.edu/IRASdocs/hires\_over.html}}.

In Section 2 of this paper we present the sample selection criteria
for the objects examined here, and present the data reduction
techniques used for reconstructing the IRAS images and measuring the
galaxy fluxes. The fluxes at each {\it IRAS} wavelength are presented
in tabular form, and contours of the infrared emission are shown
overlaid on optical images of the galaxies. Section 3 presents 
properties of the catalog, and some results derived from them. 
Appendix A presents additional notes for selected galaxy systems. 
Finally, in Appendix B we
include data for galaxy systems that were originally included in the 
BGS, but were subsequently dropped from the RBGS after a reanalysis
of their fluxes. These objects are provided for the interest of the 
reader, but do not bear on the analysis of the catalog.

\section{Data}

\subsection{Sample}

All of the targets were selected from the IRAS Revised Bright Galaxy
Sample \citep{rbgs}.  The RBGS consists of all 629 galaxies detected by
IRAS having a 60$\mu$m flux density greater than 5.24 Jy, and is thus
similar to and includes all of the well-studied Bright Galaxy Sample
\citep{bgs}, but extends coverage to the entire sky at Galactic
latitudes $|b| >$5\degree.

The following criterion was applied in order to select close pairs from the 
RBGS:

\begin{equation}
\centerline{${{2 S_{12}}\over{D_1 + D_2}} \leq 3$}
\end{equation}

\noindent where S$_{12}$ is the distance between galaxy centers and
D$_1$ and D$_2$ are their optical diameters, as measured from the
Palomar Sky Survey.  This criterion therefore
selects all systems where the galaxies are separated by less than three
times their average diameter.  Note that this excludes very advanced
mergers such as Arp 220, where the individual galactic disks can no
longer be distinguished.  This also has the additional benefit of
selecting systems that are sufficiently separated as to be resolvable
with HIRES. As such, the sample includes all of the galaxies listed in
Table 1 of \citet{hires}.  Additionally, in an attempt to resolve
sources listed in the RBGS which were likely to be confused due to
small separations, we included all small galaxy pairs with apparent
separation less than 4\arcmin.  This separation was determined by the
normal survey resolution of 4\arcmin , which in turn is set by the 
IRAS
100$\mu$m detector size.

\subsection{Data Reduction}

The IRAS data were processed in a manner similar to \citet{hires}.  
The raw detector scans were initially extracted from the IRAS database 
using the SNIPSCAN process.  These raw detector scans were then 
flattened using an iterative fitting technique that removed the 
detector baselines, and they were then deglitched in order to remove 
artifacts such as cosmic ray hits using the LAUNDR process.  The HIRES 
process was then applied to the detector scans.  Restoration was done 
on 1\sq \arcdeg \ fields in order to improve detector baseline coverage, 
with a pixel size of 15\arcsec pixel$^{-1}$, which is sufficient to 
adequately sample the restored IRAS beam.  The algorithm 
was iterated 20 times as further iterations tend only to increase 
noise amplification with little improvement in resolution.

In order to aid in the interpretation of the IRAS data, optical images
were extracted from the Digital Sky Survey (DSS) and the IRAS data
were overlaid on them (Figure 1).  This was valuable in interpreting
the correspondence between the resolved IRAS objects and the optical
galaxies.  In some cases there were small, uncataloged optical
galaxies in the DSS images, and the DSS images were used to derive
their positions.  The optical images have a pixel size of 1.7\arcsec. 
The astrometry of the optical images is based on a linear
approximation to the polynomial plate solution provided by STScI, and
yields positions accurate to roughly 2\arcsec \ \citep{laidler}.  The
astrometry of the IRAS images is limited by the pointing accuracy of
the satellite, which was approximately 2\arcsec \ in the in-scan
direction and 10\arcsec \ cross-scan \citep{irasexpsupp}. 
Furthermore, the astrometry of point-like sources produced by the
HIRES technique is known observationally to be approximately 20\arcsec
\ \citep{hires-validate}.  Thus, registration of the images should be
accurate to within one or two HIRES pixel elements, and certainly
should be better than the typical IRAS beam size.  Optical
identifications were made using the coordinates and names given in the
NASA/IPAC Extragalactic Database (NED), which are in turn derived from
the Third Reference Catalogue of Bright Galaxies (RC3, de Vaucouleurs
et al.  1991).  When no identifications were available, the galaxies
were identified directly from the DSS images and are labeled from
northeast to southwest.

Photometry was accomplished in two ways.  When the galaxies were
cleanly separated, aperture photometry was performed via the 
IPAC-Skyview software using polygonal
apertures of a size sufficient to insure that all of the galaxy flux
was measured.  In those cases where components appeared to be resolved
but not separated, the data were modeled with 2-d elliptical Gaussians
using the AIPS IMFIT and JMFIT routines.  This is justified in that
although the geometry of the IRAS beam is variable, it usually has
roughly the form of an elliptical Gaussian whose exact size and
orientation depend on the detector scan geometry.  Peak positions were
constrained to the position indicated by the nearest separated IRAS
wavelength.  If none of the IRAS data were able to supply positions,
then the Gaussian centers were constrained to the locations of the
optical peaks as given by the RC3, where possible, and otherwise
according to positions measured directly from the DSS. Note that the 
latter could introduce a bias in that it presupposes correspond
between the infrared and optical centers. In cases where the galaxies 
are well-separated at optical wavelengths, the optical and infrared 
peaks do correspond. Many of the galaxies that were decomposed using 
gaussian fitting are 
also well-separated at optical wavelengths, but are too close to one 
another to be separated by IRAS. In these cases it is reasonable to 
assume that the optical peaks will correspond to the infrared peaks. 
Only in cases of advanced mergers, such as NGC~4038, would this 
assumption break down.

Table 1 presents the measured global photometry for each of the
galaxies identified in the IRAS images from the RBGS sample.  The
supporting data in Table 1 and Table 2 were taken from the NED, and relevant notes regarding this
database are given below.  It should be noted that the 
magnitudes, morphological types, 
etc.  listed by NED are generally not on any homogeneous
system, although when possible data from NED is derived from the RC3. The table is
ordered by increasing B1950 Right Ascension of the galaxy systems, as given 
by the western-most component.  The columns are as follows:

       Column 1 - The galaxy name.  Names are given in order of preference
       from the NGC, UGC, Catalog of Galaxies and Clusters of Galaxies
       (CGCG), Morphological Catalogue of Galaxies (MCG), Markarian 
       (Mrk), and 2MASS
       catalogs.  Relevant cross-id's are also given.  The given names are
       those associated with a specific coordinate as given by the RC3, NED,
       or the Arp atlas.  The Arp name associated with a given galaxy group
       is listed with the first (westernmost) object, but no special
       significance is indicated by this.

        Column 2,3 - Equinox 1950 coordinates.  The given coordinates are
        those of the centroid of the infrared emission in cases where there
        was a separated infrared detection.  Otherwise, the optical position
        from NED is listed.  In almost all cases these coordinates originate
        from the RC3, although currently NED lists values from the Two Micron
        All Sky Survey (2MASS).  If NED listed no coordinates, then the
        coordinates are those measured directly from the DSS. Equinox B1950
        coordinates were chosen because this is the epoch of the positional
        calibration for the IRAS Level 1 Archive scans used to construct the
        HIRES images.

Column 4 - The position type. If ``O'', then the given position is an 
optical center, and if ``I'' then a new infrared center derived from 
the HIRES images.

        Column 5 - Radial velocity in km $\cdot$ s$^{-1}$. In all 
        cases where these are listed they are spectroscopic redshifts 
        from a variety of sources as given by NED. Uncertainties are 
        typically 10-100 km/sec.

        Column 6 - Total optical magnitude as given by NED. In most cases
        this is the blue magnitude listed in the RC3.

        Column 7-14 - 12, 25, 60, and 100 $\mu$m integrated fluxes in Jy, and
        the associated 1-$\sigma$ \ uncertainties. 
In cases where the targets were also given by \citet{hires}, the 
objects were remeasured in order to ensure uniformity of calibration 
with the rest of the survey (see Section 2.3).
In some
cases the galaxies were still unresolved at all wavelengths. In this 
case only the global flux, as measured from the HIRES data, is given.
In cases where one or more components remained unresolved, but some
components were resolved, the brightest unresolved component in the 
resolved waveband gives the flux
of all the components, and subsequent unresolved components are marked
with ellipses.  Upper limits are denoted with a ``$<$'' and are the 
flux measured at the known optical location of the galaxy in an 
aperture that has the same size as the effective IRAS beam.

        Column 15 - Log of the far-infrared luminosity, in units of 
solar luminosities. This is the luminosity from 40---122$\mu$m 
 \citep{helou}. H$_o$ = 75 km s$^{-1}$ mpc$^{-1}$ is assumed.  This 
 quantity is useful for photometric study,is not very sensitive 
 to the shape of the spectral energy distribution, and is the 
 quantity tabulated by \citet{hires}.
The L$_{fir}$ described in the RBGS is the luminosity from 1---500$\mu$m. It 
was derived by applying a correction factor to the flux between 
40-122$\mu$m based on the 60/100$\mu$m 
color (Lonsdale et al. 1985). For the galaxies described here, the 
median correction factor is 1.44$\pm$0.09. In other words, Log L$_{1-500\mu 
m}$ = Log L$_{40-122\mu m}$ + 0.16. This is also different from the 
quantity L$_{ir}$ as described by \citet{araa}, which is the flux from 
8---1000$\mu$m, but which generally cannot be computed here since it 
requires detections in all four IRAS bands.

\subsection{Photometric Uncertainties}

Evaluating the photometric uncertainty of the HIRES data product is 
quite difficult.  In general, uncertainties arise from three sources, 
all of which vary in importance depending on the particular field.  
The integrated flux density uncertainties quoted in Tables 1 \& 2 
contain measurement and confusion errors, but {\it not} systematic 
effects in the overall calibration.

Confusion is the first limitation.  The dominant source of noise in 
HIRES is not photometric background noise, but confusion due to noise 
spike amplification.  High sigma noise peaks are amplified by the 
deconvolution process; they appear similar to weak point sources with 
a signal strength as high as 0.1 Jy. These spikes are illustrated in 
Figure 2. This results in a highly 
non-Gaussian single-sided noise distribution on spatial scales similar 
to the beam size, not the pixel size.
In those cases where the galaxy fluxes
are less than 0.3 Jy, it becomes difficult to differentiate the target 
from amplified noise.  As a result, quoted upper limits are often 
quite high, as this upper limit is set by the flux contained in these
noise peaks.  Our achieved sensitivity is thus around 0.25-0.3 Jy, 
depending on the wavelength and field geometry. Similarly quoted 
uncertainties are often also high, depending on the amplitude of these 
spikes. Apertures similar to the effective beam size were used to 
evaluate a median false signal due to the noise spikes. These spikes 
are the dominant source of uncertainty for faint objects.

The technique used to derive the photometry is the second contributor 
to the photometric uncertainty.  In cases where the galaxies are well 
separated and aperture photometry could be used, this typically 
contributes only a few percent to the total error. 
In those cases where the galaxies were 
not well separated and Gaussian fitting was used, this becomes the 
dominant source of error and can range anywhere from 20-50\% depending 
on the degree of resolution of the targets.

Absolute photometric calibration is the third major source of
uncertainty, and which is not included in Table 1.  As noted in
\citet{hires}, there are certain caveats to the photometric
calibration of the HIRES data product.  In particular, the calibration
of the IRAS data is in part dependent on factors due to detector
responsivity and dwell time.  Known as the AC/DC correction, it is the
difference in responsivity for point sources versus extremely extended
sources, which was characterized as a function of detector dwell time
based on the nominal survey slew speed.  This is well known for point
sources, and hence the Point Source Catalog (PSC) is properly
calibrated \citep{irasexpsupp}.  However, it is slightly different for
small extended sources, and is a poorly understood function of source
extension.  This was seen during the data analysis presented in Paper
I, where it was found that the majority of the HIRES fluxes were
significantly greater than the values estimated using one-dimensional
coaddition with the ADDSCAN/SCANPI processing available at IPAC. In
Paper I this issue was wholly avoided by forcing all of the data onto
the same flux scale as the PSC by using the component flux ratios
indicated by HIRES to divide up the flux indicated by the BGS, which
was produced using the ADDSCAN process which is known to have the same
photometric scale as the PSC.

Figure 3 through 6 show the difference in flux estimates between this
paper and the ADDSCAN/SCANPI values published in the RBGS (Sanders et 
al. 2003). The data points shown in these figures are limited
to cases where the flux referred to in the RBGS was unambiguous. 
These are primarily systems which were either unresolved by HIRES (and
hence both catalogs have single fluxes) or were sufficiently separated
as to have been resolved by both ADDSCAN/SCANPI and HIRES (e.g. NGC 875). 
The mean offsets between the two catalogs are 27.5, 12.8, 4.5, and 5.5 \%
at 12, 25, 60, and 100 \micron .  The observed scatter around the mean is similar 
to the estimated flux uncertainties in Table 1. These are particularly
significant in the faint 12\micron \ channel, where although the 
miniumum requirement for reporting is a S/N$>$3, the average detection 
only has a S/N$\approx$5.  The mean offsets are also 
similar in size to the observed scatter. Testing of
HIRES has shown that measured integrated fluxes of 
unresolved point sources have an intrinsic scatter of about 8-12\%
compared to those of the PSC \citep{hires-validate}.  In all cases the
one sigma scattter in offsets observed for the galaxies is larger than
the value of the systematic offset.  There is no statistically
significant trend as a function of flux.  In several cases the
statistically significant outliers seen in the brighter channels are a
result of differences in background estimation between the
one-dimensional ADDSCAN results and the two-dimensional HIRES results.

While the version of HIRES used in this paper produces data believed
to be on the AC scale, appropriate for point sources, it is clear that
there are systematic offsets relative to other AC-calibrated IRAS data
products.  Previous experiments in Paper I showed that the HIRES data
product correctly reproduces the photometry of point sources in
accordance with the PSC. As the exact source of this offset remains
unclear, as does the calibration for small extended sources, this data
has not been forced to agree with the RBGS, unlike \citet{hires}. 
This is the source of the variations between the fluxes in Paper I and
this work.

As a result of this offset, the results presented here differ from 
those based solely on PSC-calibrated products by small 
amounts. The ratio of 60 to 100\micron \ flux remains unchanged, as the 
offset is the same in both bands. The infrared luminosities are 5\% 
higher, a value considerably less than the typical uncertainty. The 
log of the 12 to 24\micron \ ratio differs by being 0.05  higher. When 
appropriate, these offsets will be discussed in the Section 3 of 
this paper.

\section{Results}

\subsection{Far-IR Properties}

The cumulative distribution functions (CDFs) of {\it L}$_{\rm fir}$,
Log($f_{12}/f_{25}$), and Log($f_{60}/f_{100}$) are given in Figures
7, 8, and 9.  These distributions only include the galaxies actually
detected and resolved by HIRES. These distributions are nearly
identical to \citet{hires}.  This is to be expected, since the shape
of the CDF remains unchanged so long as
the nature of the incompleteness in the data is random.  Since Paper I
differed from this paper in being drawn from a parent sample which was
different from the RBGS primarily in spatial extent on the sky, the
CDFs are expected to remain the same.

As in \citet{hires}, a comparison sample of galaxies was constructed by 
selecting a subsample drawn from
the BGS which had no visible signs of interaction and were not in close 
pairs (this is the same sample described in Paper I).
From these isolated BGS galaxies we selected a subsample so as to have the same 
distribution of blue magnitudes as the RBGS close pairs. We can therefore 
compare the far-infrared properties of a far-infrared flux-limited sample of 
interacting pairs to a similarly flux-limited sample of isolated galaxies with the 
same distribution of optical luminosities.

The CDF for Log L$_{\rm fir}$, which is computed from the 60 and
100\micron \ fluxes, is shown in Figure 7.  The median value of {\it
L}$_{\rm fir}$ is 10$^{10.50}$ {\it L}$_{\sun}$ for individual,
resolved galaxies in the paired and multiple RBGS systems studied
here.  This is somewhat higher than found in Paper I ({\it L}$_{\rm
fir}$=10$^{10.30}$ {\it L}$_{\sun}$), and cannot be readily attributed
to the offsets in calibration, which are of order 5\% at these
wavelengths.  Kolmogorov-Smirnov statistics indicate that the isolated
and paired samples are not draw from the same parent sample at better
than the 99.99\% confidence level.  In separated galaxy pairs, then,
the interaction process enhances {\it L}$_{\rm fir}$ by a factor of
roughly 3.

Similar differences are seen in the far-IR colors.  The median value 
of 
Log($f_{12}/f_{25}$) is -0.36 for resolved component galaxies in pairs vs
-0.19 for isolated galaxies.  Paper I found -0.43 and -0.2.  However,
as noted earlier, this may be a result of the differing flux
calibration between Paper I and this paper.  Adjusting for this
produces a mean Log($f_{12}/f_{25}$) of -0.41 for individual galaxies
in the RBGS HIRES interacting galaxy sample.  The maximum difference
in the CDF is 0.31 and occurs at Log($f_{12}/f_{25}$) = -0.28. 
Kolmogorov-Smirnov statistics reject the null hypothesis that these
two sample are drawn from the same parent sample at better than the
99.99\% level.

A median Log (f$_{60}$/f$_{100}$) value of -0.25 is observed for galaxies
belonging to pairs and groups in the RBGS HIRES sample, compared to
-0.34 for non-interacting RBGS galaxies.  This is the same result as
seen in Paper I. The difference in CDF between the two samples is less
pronounced overall than at the shorter wavelengths.  Nevertheless the
maximum difference in CDFs is 0.34 at Log($f_{60}/f_{100}$)=-0.29. 
Again, we can reject the null hypothesis that the two samples were
drawn from tha same sample at better than the 99.99\% level.

\subsection{Pairing in the Far-IR}

While there is clearly evidence that pairs and groups of galaxies
generally have higher star formation activity compared to isolated
galaxies, there is still uncertainty regarding the relative degree to
which enhanced star formation is triggered in individual galaxies
during various phases of the interaction and merger process.  Naively,
among spiral-spiral pairs that have companions with nearly equal B-band
luminosities, one would expect that both galaxies contribute in similar
proportions to the total far-infrared emission of the pair.  Previous
authors, as discussed earlier, have generally concluded that for very
distant pairs observed in the far-IR only one galaxy is infrared
active.  Other authors, working at optical and near-IR wavelengths,
have reached the same conclusion using indirect measures of star
formation. Using the higher resolution images presented here, it is 
possible to test this result over a much wider range and smaller 
absolute separations than previously possible.
Figure 10 plots the measured flux
ratios (solid circles) and upper limits (open circles) at the longest
resolvable wavelength between the brightest galaxy and its companion as
a function of the total far-infrared luminosity, {\it L}$_{\rm fir}$.
For galaxy groups, the ratio plotted is the flux of the brightest
galaxy divided by the average flux of the companion galaxies in the
group.  These results provide little evidence that both companions
contribute comparably to the infrared emission, and that there is no 
increased tendency for infrared luminous galaxies to be found with 
other infrared luminous galaxies.  That is, 66\% of the
interacting systems have component flux ratios greater than 3, 56\%
have ratios above 5, and 36\% have ratios greater than 10.  
\citet{hires} claimed that in approximately 2/3 of interacting pairs, the
ratio of the flux densities of the companions are less than 10, which
is confirmed here in the much larger sample of infrared-bright galaxy
system investigated here.

Figure 10 also shows that over the range of {\it L}$_{\rm fir}$ spanned by this
RBGS subsample, there is no clear correlation between the companion
galaxy flux ratios and {\it L}$_{\rm fir}$.  Although at flux levels
Log(L$_{fir}$/{\hbox{{\it L}$_\odot$\ }}) $>$ 11 there suggestively are 
almost no systems where the component
flux ratio is less than three, our sample does not extend to such high 
luminosities as would produce a clearer result.  This is because our selection criteria biases us against
very advanced mergers, which are the majority of the luminous and
ultraluminous infrared galaxies.  Since both components in an
interacting system are presumably undergoing a similar degree of tidal
disruption, it seems that in the relatively early stages of interaction
sampled here, the details of the encounter itself are less important
than characteristics of the individual galaxies in determining the
degree of far-infrared enhancement.  Major factors expected to play an
important role are the molecular gas content available to fuel star
formation and the mass of the stellar bulge which may regulate the
degree of accretion onto a supermassive black hole.

Finally, we can examine whether or not confusion has caused 
galaxies to be erroneously included in the RBGS because their 
combined flux was high enough to meet the flux limit criterion, 
but which would not have been selected if they could be resolved.
Figure 11 presents the cumulative distribution functions of 60$\mu$m
fluxes for the HIRES-resolved galaxies above and below the 5 Jy
selection limit.  An examination of Table 1 shows that there are 2
systems that appeared in the RBGS by virtue of having a combined,
unresolved flux above the 5.24 Jy limit at 60$\micron$ but whose
individual components were all clearly below this limit (IC563/4, NGC
7752/3).  An additional five (IC 2522/3, UGC 6436a/b, NGC 3991/4/5, MCG
-03-34-063a/b, and VV 414) have only one component whose flux including
uncertainties may be as high as the RBGS flux limit.  Thus, close pairs
in the RBGS verifiably affect the selection of the sample at only the
0.3\% level (2/629), and at worst may account for 1.1\%.  This
statement applies to separated pairs resolved by HIRES in this study.
Very close pairs with separations less than $\approx$30\arcsec \
(typically ongoing mergers in the local Universe) which cannot be
resolved by HIRES account for increasingly larger fractions of RBGS
objects as a function of increasing total far-IR luminosity (e.g., see
review by Sanders \& Mirabel 1996).  The manner in which the total
far-IR fluxes of such objects are distributed between the individual
components remains unknown for such pairs, and these will be fruitful
targets to study with higher resolution using observatories such as
SIRTF and SOFIA.

\subsection{Optical Morphology and far-IR Enhancement}

Recent computational models by \citet{mihos1,mihos2} have predicted 
that the presence of a large central bulge in a galaxy will help 
stabilize it against tidal perturbation.  Specifically, they found 
that in major mergers of galaxies the presence of a central bulge 
inhibits the flow of gas into the central few kiloparsecs of a galaxy, 
thus preventing high gas densities from being quickly reached and 
suppressing any period of rapid star formation until the end of the 
merger \citep{mihos2}.  Thus, late-type spirals are expected to 
experience starbursts during the initial stages of merger, while 
early-type spirals undergo strong starburst activity only during the 
completion of the merger process.  This provides a mechanism to delay 
the onset of starburst activity in some systems until very advanced 
merger stages are reached, as otherwise it is difficult to invoke a 
starburst model for ultraluminous infrared galaxies (which appear to 
be very advanced mergers) given the expected timescale for starbursts.  
Since very evolved mergers have such disturbed morphologies that it is 
difficult to determine the form of the merger progenitors, the most 
viable observational test is to examine young merger systems which 
have not evolved as far away from their original forms and look for 
evidence for the onset of enhanced far-IR activity in bulgeless 
galaxies.

Hubble types were taken from NED. A fraction of the galaxies either
have not been classified at all, or are simply listed with generic
types such as ``spiral''.  All other resolved spiral galaxies that
were actually classified were considered to be either ``early type''
(S0, SB0 through Sa, SBa) or ``late-type'' (Sb,SBb and higher). 
Although the specific Hubble type for each galaxy was kept track of,
for purposes of this analysis it was felt to be more useful to group
the types into such very broad categories in order to improve the
counting statistics.

Only 11 ellipticals known from optical imaging of the galaxy pairs are
found in the entire sample.  Of these, none are detected at both 60 and
100$\mu$m and only three are detected by IRAS at any wavelength.  These
numbers agree with what would be expected based on a random pairing of
elliptical and spiral galaxies given an elliptical/spiral fraction
similar to that of field galaxies, or are perhaps a little low.  In
particular, this is the number expected if we assume that every system
contains one bright spiral galaxy, and that the remaining faint
galaxies are distributed according to the field elliptical/spiral
ratio.  This also agrees with the low detection fractions for
elliptical galaxies found by other studies \citep{sulentic,haynes}.

An examination of the late and early-type spirals in the sample
indicates that there are no differences between these populations.
The rate of detection for both of the two classes is around 88\%, 
indicating that they have a similar fraction of their distribution 
above our detection limit.
This is further illustrated by Figure 12, which shows the cumulative 
distribution function of Log {\it L}$_{\rm fir}$ for the galaxies which were 
actually detected in the two classes. They are extremely similar. The 
K-S test cannot reject the null hypothesis that the two samples are 
drawn from tha same sample with better than 65\% confidence.
Results are the same for the color ratios Log($f_{12}/f_{25}$) and 
Log($f_{60}/f_{100}$).

Similar results also hold for different combinations of Hubble
sub-types, such as considering only S0 galaxies as ``early''. 
\citet{haynes} found that the detection rate for isolated spiral
galaxies is roughly independent of morphological type. 
\citet{roberts} have confirmed this using large optical samples of
isolated galaxies and comparing them against the IRAS data.  The
median, 25\% and 75\% percentile values of the distribution of {\it
L}$_{\rm fir}$ for the sample spiral galaxies are also very similar to
those found by Roberts \& Haynes for isolated UGC galaxies.  This is
unsurprising, since both the results presented here and previous
studies have shown that only a small increase in {\it L}$_{\rm fir}$
occurs for widely separated (non- overlapping) interacting pairs
\citep{haynes,hires}, and therefore any enhanced far-IR activity that
could distinguish late from early spirals is likely to be slight. 
This enhancement would be further diluted by the presence of systems
that have not yet reached first perigalcticon, and hence have not yet
reached the point at which the two classes would separate themselves
\citep{mihos2}, and by the presence of systems which are unlikely to
actually merge or otherwise strongly interact with each other.

\section{Conclusions}

We have presented an atlas of high resolution IRAS observations of 
all 106 of the paired (and in many cases interacting systems) in the 
IRAS
Revised Bright Galaxy Sample  with a 
60$\mu$m flux density greater than 5.24 Jy. The atlas contains 
infrared contours overlaid on optical images and a catalog of fluxes 
or upper limits in all four IRAS wavebands.

We have presented the infrared luminosities and colors of the paired 
galaxy sample, and compared them to a sample of isolated galaxies. We 
find substantially the same results as \citet{hires}, namely that the 
paired galaxies have a measurably different distribution of infrared 
properties than isolated galaxies.

Using morphological optical classifications for the galaxies
we conclude that there is no difference between late and early-type 
spirals in terms of their far-IR properties.  In particular, no 
significant enhancement is seen in the far-infrared luminosity or 
color of the late-type spirals as compared to the early-type spirals.  

\acknowledgements

We would like to thank Ron Beck and John Fowler for their assistance 
in setting up our HIRES processing facility, and in particular Diane 
Engler for her speedy work in extracting the detector scans from the 
IRAS database. We thank an anonymous referee for his or her comments 
that led to an improved presentation of these results.

This work has made use of the NASA/IPAC Extragalactic Database (NED)
which is operated by the Jet Propulsion Laboratory, California
Institute of Technology, under contract with the National Aeronautics
and Space Administration.  It has also made use of the Digitized Sky
Survey, produced at the Space Telescope Science Institute under U.S.
Government grant NAG W-2166.  This was based on photographic data
obtained using the Oschin Schmidt Telescope on Palomar Mountain,
operated by the California Institute of Technology, and the UK Schmidt
Telescope, which was operated by the Royal Observatory Edinburgh and
the Anglo-Australian Observatory.  JAS and JMM were supported by the
Jet Propulsion Laboratory, California Institute of Technology, under
contract with NASA. DBS acknowledges support from a Senior Award
from the Alexander von Humboldt Foundation and from the
Max-Planck-Institut fur extraterrestrische Physik as well as support
from NASA grant NAG 90-1217.

\appendix

\section{Notes on Individual Galaxy Systems}

\noindent{\it NGC 520} --- the flux in this advanced merger appears to 
be centered between the two galaxies, in what appears to be a dust lane.

\noindent{\it IC 2163} --- this system is somewhat puzzling in that
there is significant emission to the west of the center of the
western galaxy in the pair. Additionally, in general the peak flux
appears to occur between the galaxies. It is notable that the
reconstructed {\it IRAS} beam is unfortunately elongated at nearly
the same position angle as the two galaxies, and hence the irregular 
beam is confusing the location of the emission.

\noindent{\it NGC 4038/9} --- the emission originates between
the galaxy centers in the region where the disks overlap. This is
consistent with the results of \cite{vigroux}.

\noindent{IC 4153} --- there is a sizable discrepancy between the 
HIRES flux and the ADDSCAN flux used by the RBGS at 100 \micron .  
Furthermore, the HIRES flux given in Table 1 agrees with the flux 
indicated by the FRESCO data product. FRESCO is a two-dimensional 
coadd data product available from IPAC. This coadd is not an 
iterative reconstructed image, and hence should not suffer from 
iterative artifact amplification. An examination fo the complex 
structure seen in HIRES and FRESCO images surrounding this source, as 
well as the details of the ADDSCAN processing, show that this 
discrepancy is probably due to differences in the baseline 
(background) fitting.

\noindent{\it NGC 5953/4} --- given as CPG 468 by \cite{donovan}. The 
HIRES data agrees with the higher resolution ISOPHOT data indicating 
the dominance of the southwestern component.

\noindent{\it NGC 6907/8} --- the galaxy NGC 6908 is actually a small 
spiral galaxy
superimposed on the NE arm of NGC 6907, and is most clearly seen in 
near-infrared images. It cannot be resolved by HIRES. This system appears in Appendix B.

\noindent{\it NGC 7752/3} --- also known as CPG 591.  \citep{donovan}  
find that the southwestern component is more peaked and dominates the 
ISOPHOT 60 and 100\micron \ data. The HIRES data supports this finding, 
particularly in the mid-IR. 
%Results from \citet{xugao} indicate 

\section{Additional Galaxy Systems}

In addition to the 106 galaxy systems detailed in Section 2, several 
other systems were also processed. During the compilation of the
RBGS \citet{rbgs}, the IRAS data were recalibrated, and the choice of flux measures
used to estimate the IRAS fluxes changed. As a result, there are a
handful of systems which appear in the BGS$_{1}$+BGS$_{2}$ but not
the RBGS, and vice versa. These are detailed in the RBGS, Section 3.2.
As a result, some additional galaxy systems were processed with
HIRES, but do not belong in the RBGS sample proper and were not
included in Table 1 of this paper. They are presented here for 
informational purposes.  The images are shown in Figure 13, and the 
tabulated fluxes appear in Table 2.

\clearpage

\figcaption{HIRES data for each pair or group in the RBGS overlaid on
gray-scale images from the Palomar Digital Sky Survey. The contours are
at intervals of 10, 16, 25, 40, and 63\% of the peak flux; they show the
results of 20 iterations of the MCM algorithm, and represent the highest
achievable IRAS resolution at 12, 25, 60 and 100$\mu$m. The axis labels are
B1950 coordinates. The scale bar inside each 100$\mu$m panel represents 50 kpc
or 25 kpc at the distance of the system (as labeled).
  }

\figcaption{Illustration of amplification of high-sigma noise outliers
by the HIRES process, resulting in noise ``spikes'' similar in size to
the reconstructed IRAS beam and having total fluxes of $\approx$ 0.1
Jy.  This is the 60 $\mu$m image of Arp 271.}

\figcaption{Fractional difference between one-dimensional 
ADDSCAN/SCANPI fluxes \citep{rbgs} and HIRES fluxes at 12 \micron . 
The horizontal axis is the RBGS flux in Jy. The vertical axis scaling is such 
that 0.2 indicates a difference of 20\%.}

\figcaption{Same as Figure 3, but for 25 \micron .}

\figcaption{Same as Figure 3, but for 60 \micron .}

\figcaption{Same as Figure 3, but for 100 \micron .}

\figcaption{{\it L}$_{\rm fir}$ cumulative distribution function for 
single, isolated galaxies in the RBGS
and individual galaxies in multiple systems resolved by HIRES. The 
resolved individual galaxies in RBGS paired systems show a small 
($\approx$3$\times$) increase in {\it L}$_{\rm fir}$ compared to 
isolated galaxies}

\figcaption{CDFs of Log(f$_{12}$/f$_{25}$) for paired and isolated 
galaxies, as in Figure 7.}

% figure 9
\figcaption{CDFs of Log(f$_{60}$/f$_{100}$) for paired and isolated 
galaxies, as in Figure 7.}

% figure 10
\figcaption{Log {\it L}$_{\rm fir}$ versus the pair component flux ratio. 
Closed circles are measured ratios, while open circles are upper limits.
No correlation is seen. The distribution of lower limits is  
somewhat random, since there was a fixed lower flux limit for 
the dim component, while the bright component could span a wide range of 
detected brightnesses.}
%figure 11
\figcaption[CDF of 60$\mu$m flux distribution in the RBGS]{Integrated, spatially 
resolved 60 $\mu$m flux cumulative distribution functions for galaxies 
detected in the RBGS.}

%figure 12
\figcaption{Cumulative distribution function of {\it L}$_{\rm fir}$ of late and 
early-type interacting spiral galaxies in the RBGS. There is no measurable 
difference between the two.}

%figure 13
\figcaption{HIRES data as in Figure 1 for additional galaxy systems 
processed but not members of the final RBGS, as described in Appendix B.}

\clearpage

\begin{figure}
    \epsscale{0.9}
    \plotone{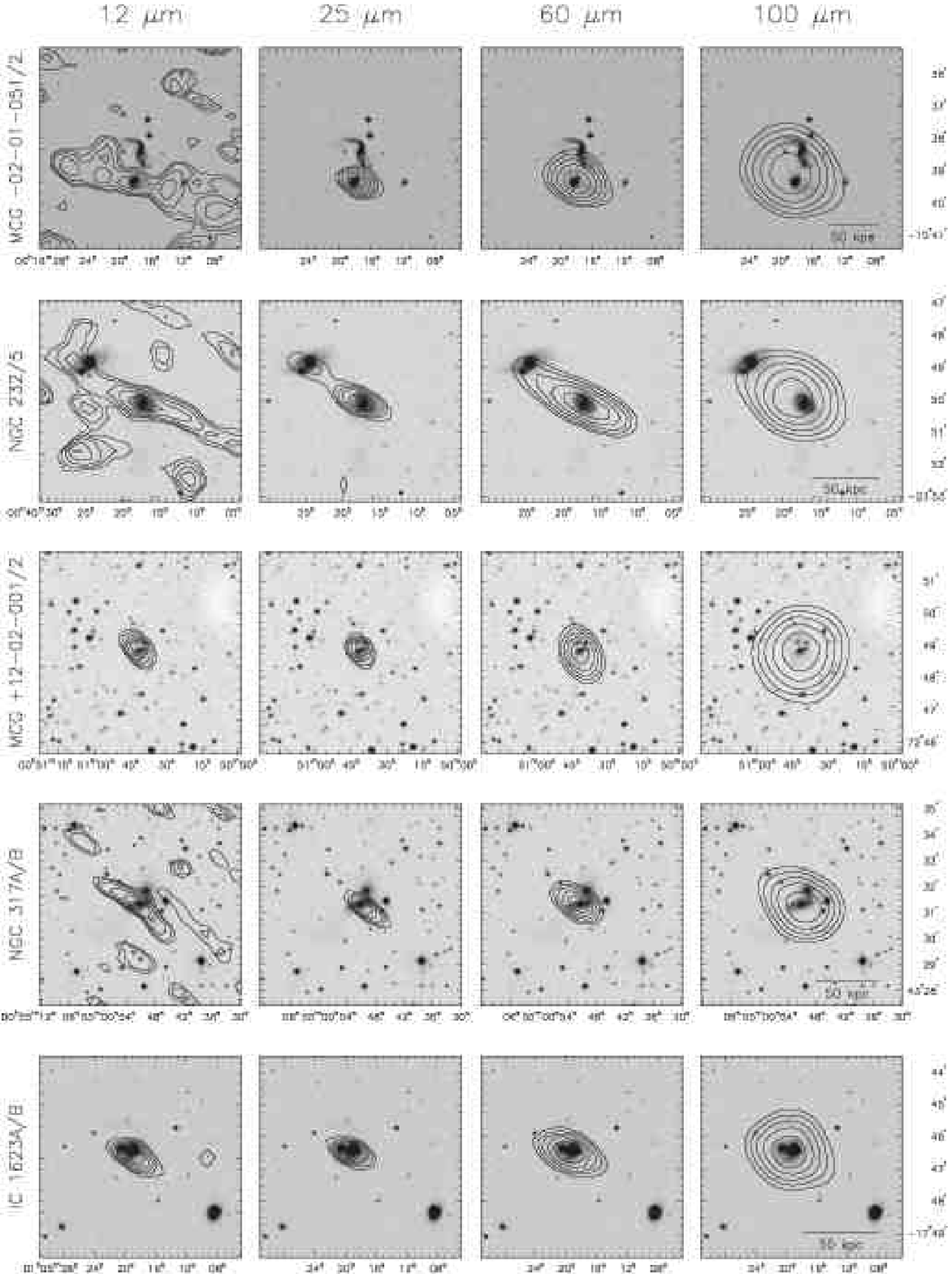}
            \caption{}
    \end{figure}
\clearpage  

\begin{figure}
    \epsscale{0.9} 
    \figurenum{1 cont}
    \plotone{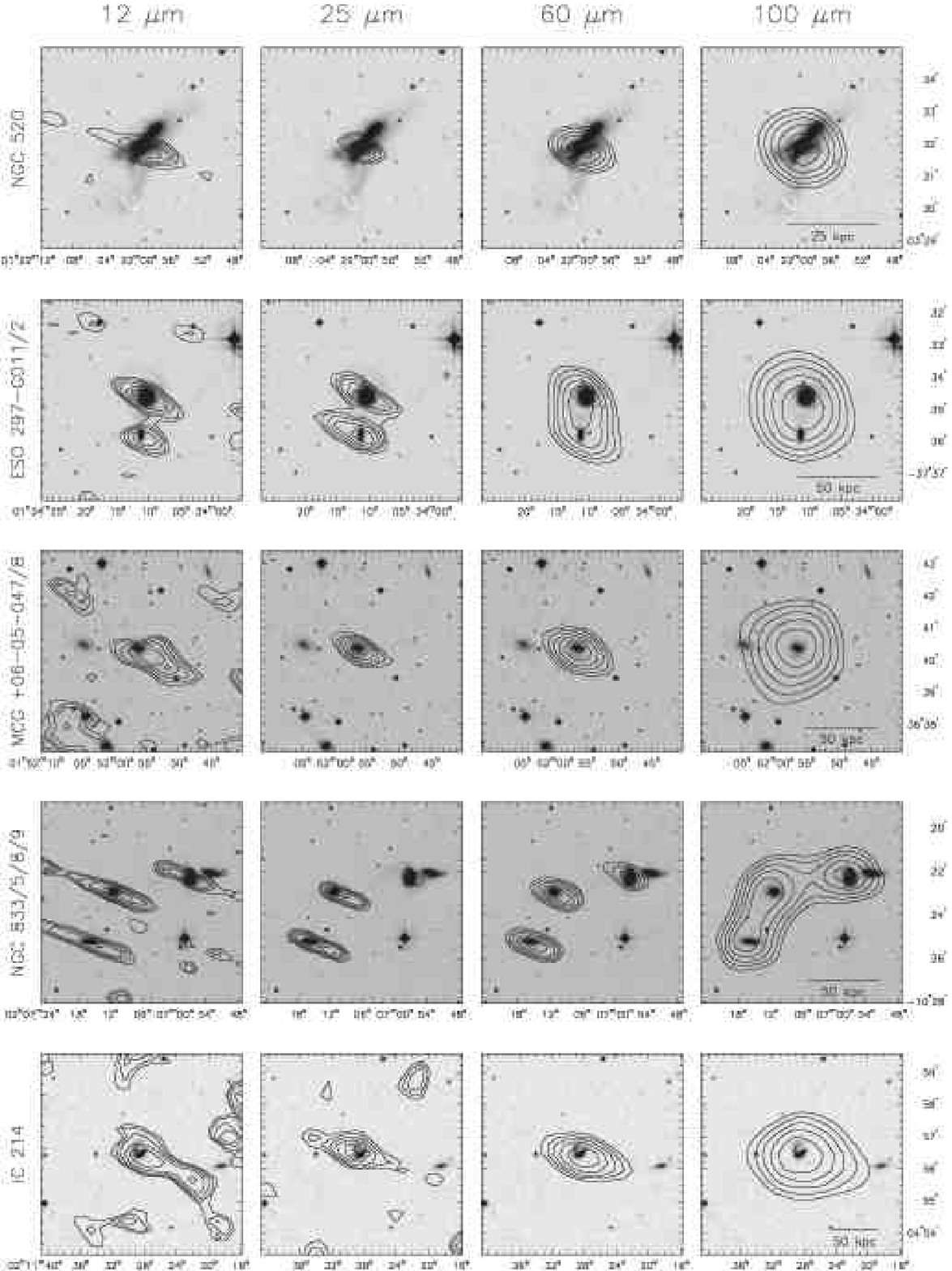}
            \caption{}
    \end{figure}
\clearpage  

\begin{figure}
    \epsscale{0.9}
    \figurenum{1 cont}
    \plotone{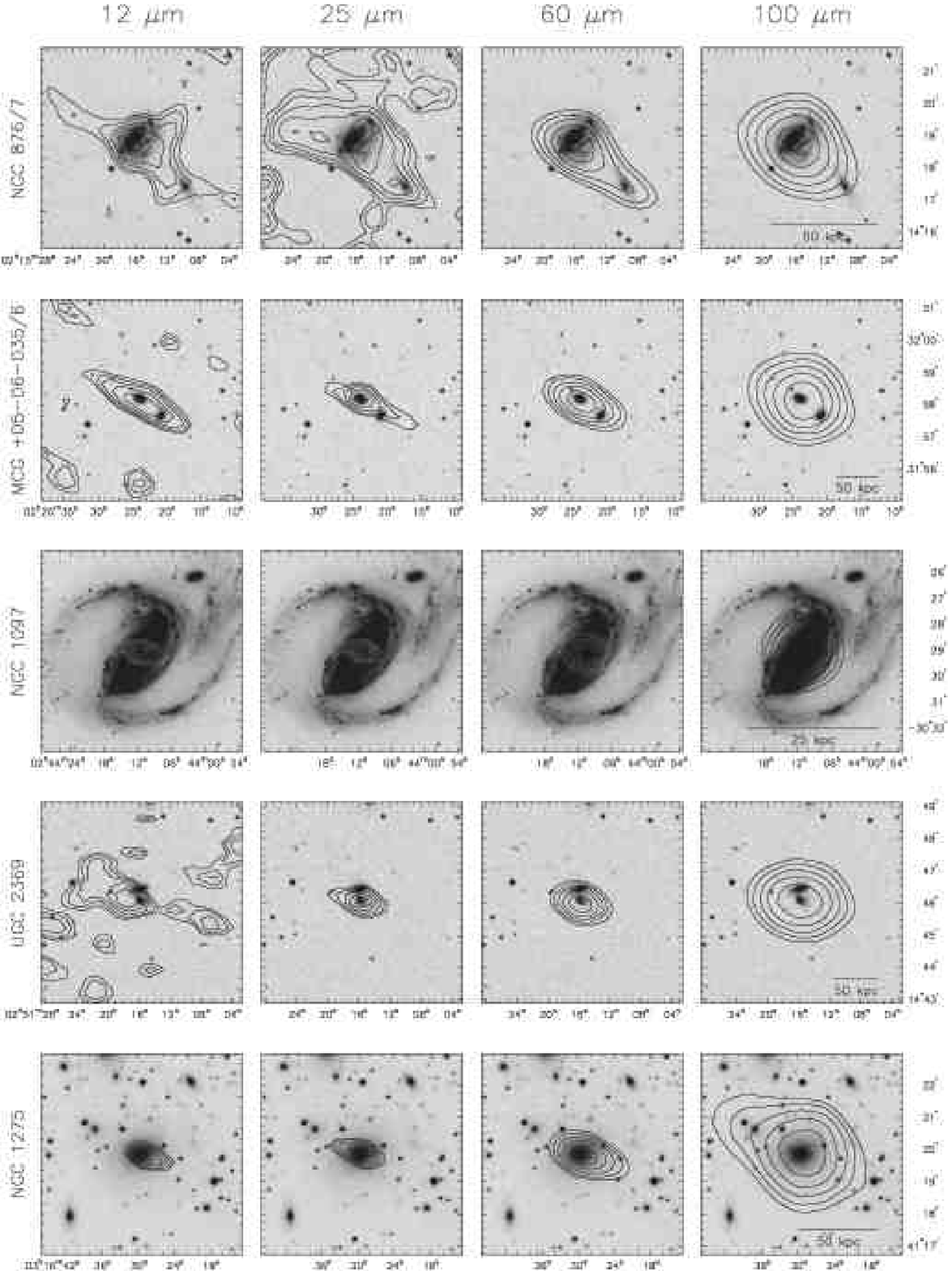}
            \caption{}
    \end{figure}
\clearpage  

\begin{figure}
    \epsscale{0.9}
    \figurenum{1 cont}
    \plotone{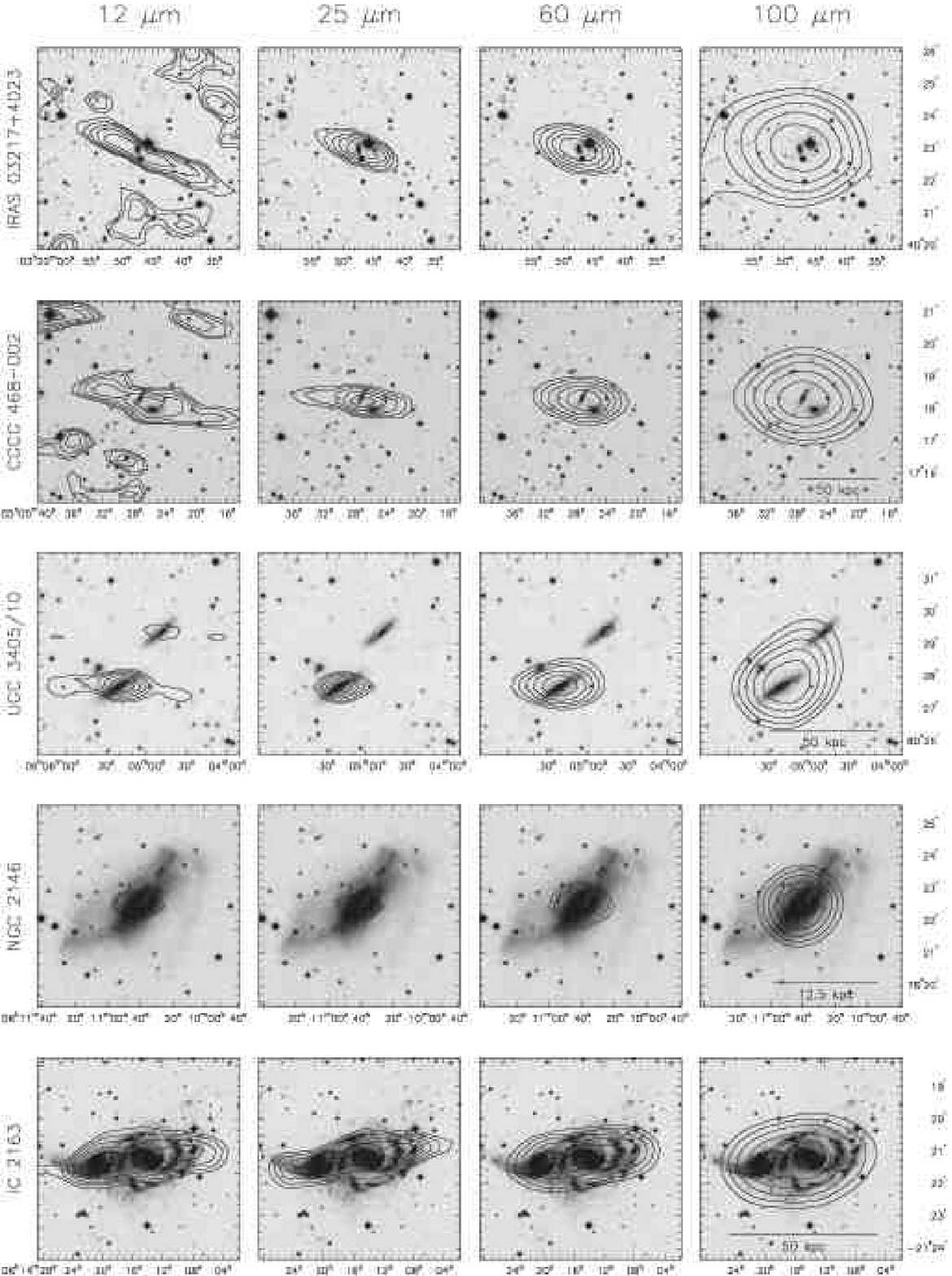}
            \caption{}
    \end{figure}
\clearpage  

\begin{figure}
    \epsscale{0.9}
    \figurenum{1 cont}
    \plotone{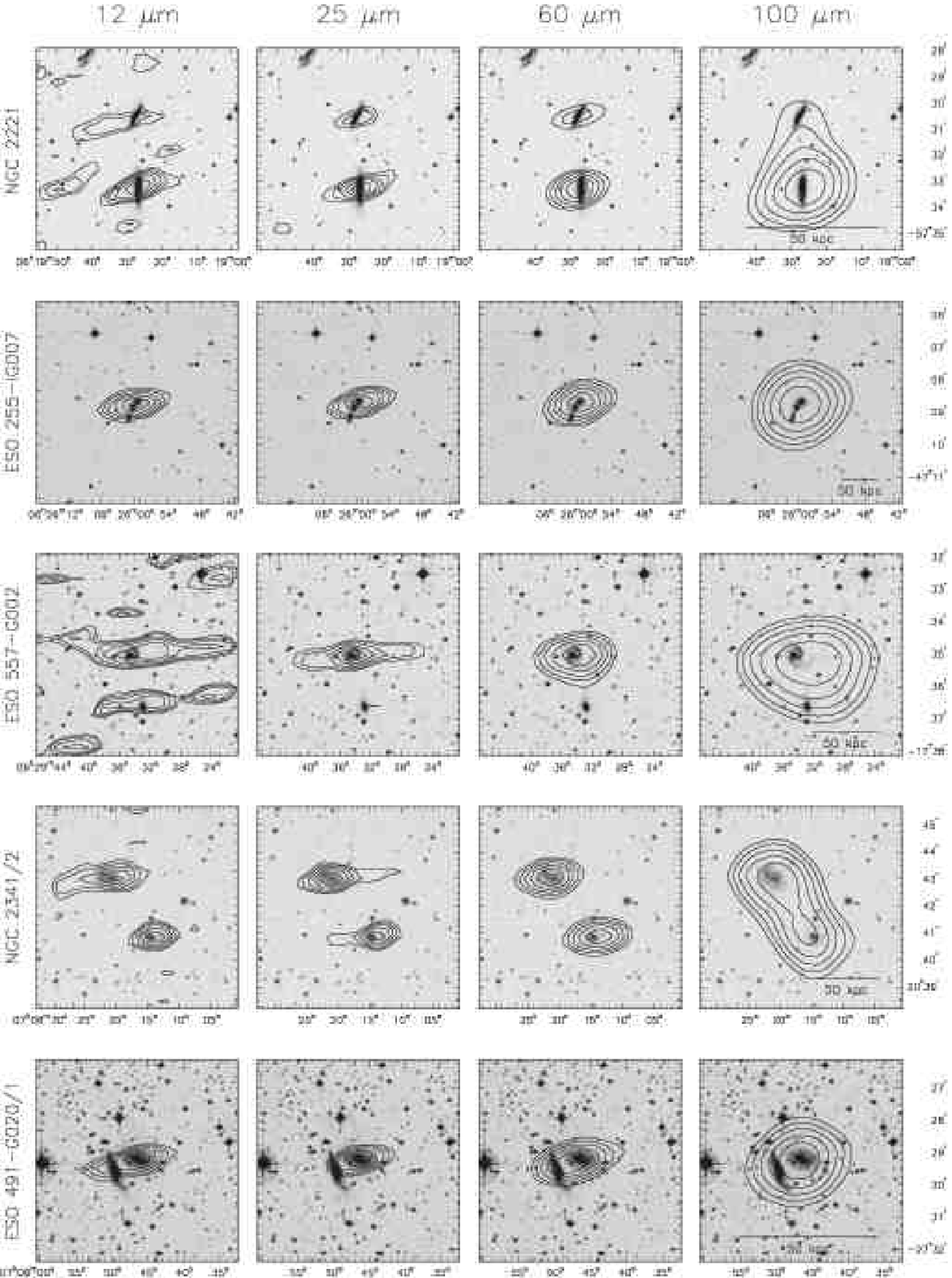}
            \caption{}
    \end{figure}
\clearpage  

\begin{figure}
    \epsscale{0.9}
    \figurenum{1 cont}
    \plotone{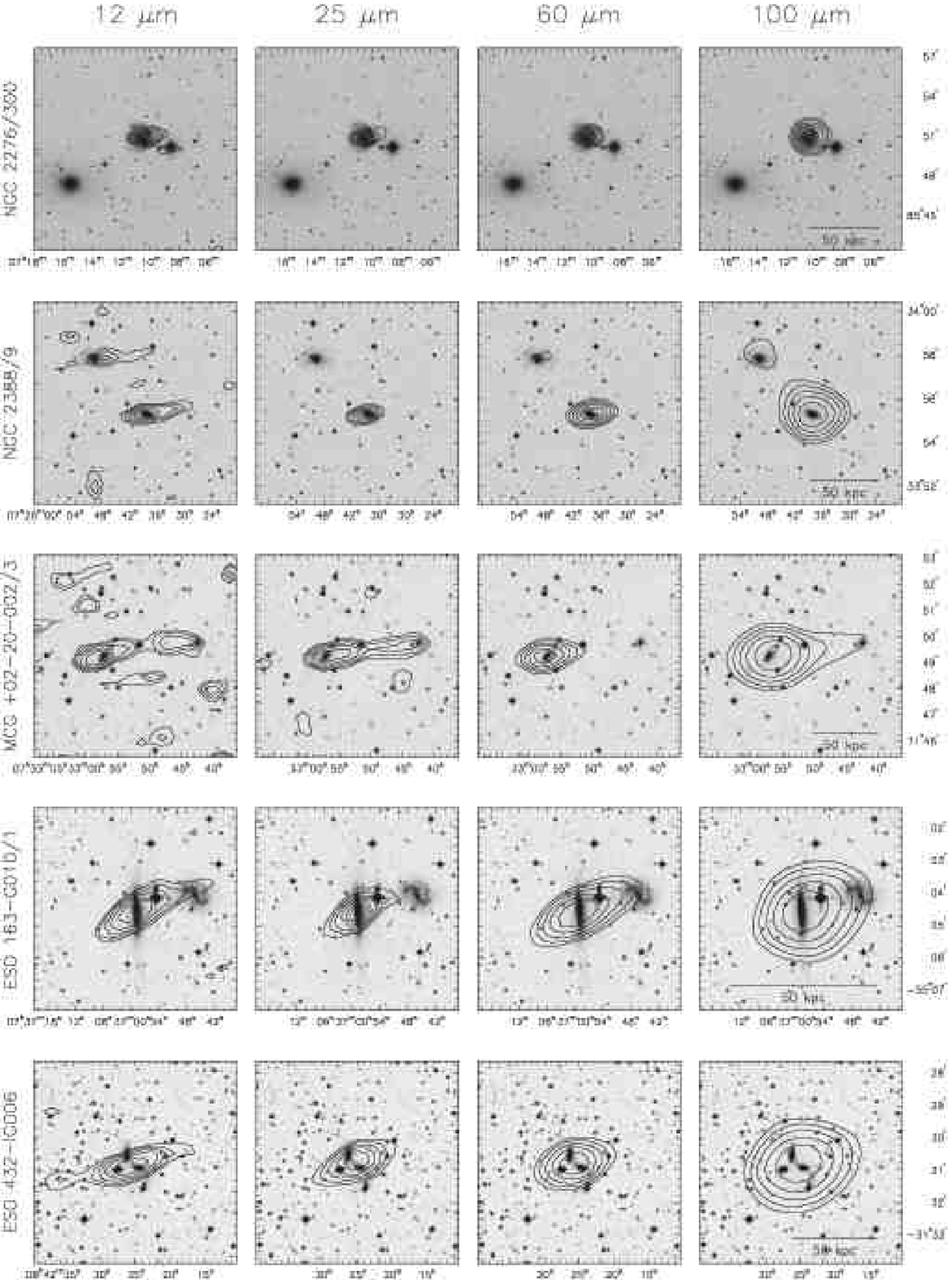}
            \caption{}
    \end{figure}
\clearpage  

\begin{figure}
    \epsscale{0.9}
    \figurenum{1 cont}
    \plotone{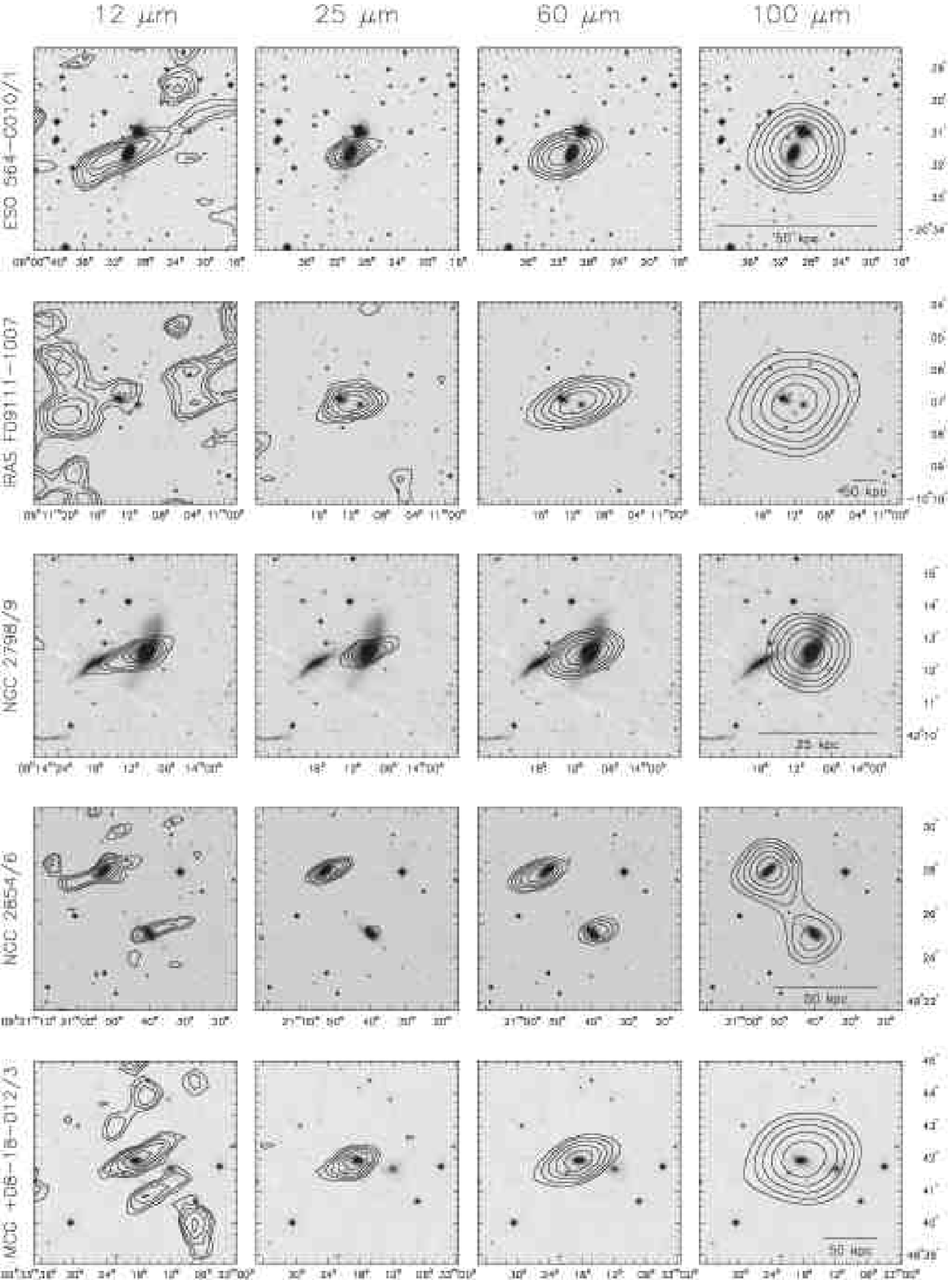}
            \caption{}
    \end{figure}
\clearpage  

\begin{figure}
    \epsscale{0.9}
    \figurenum{1 cont}
    \plotone{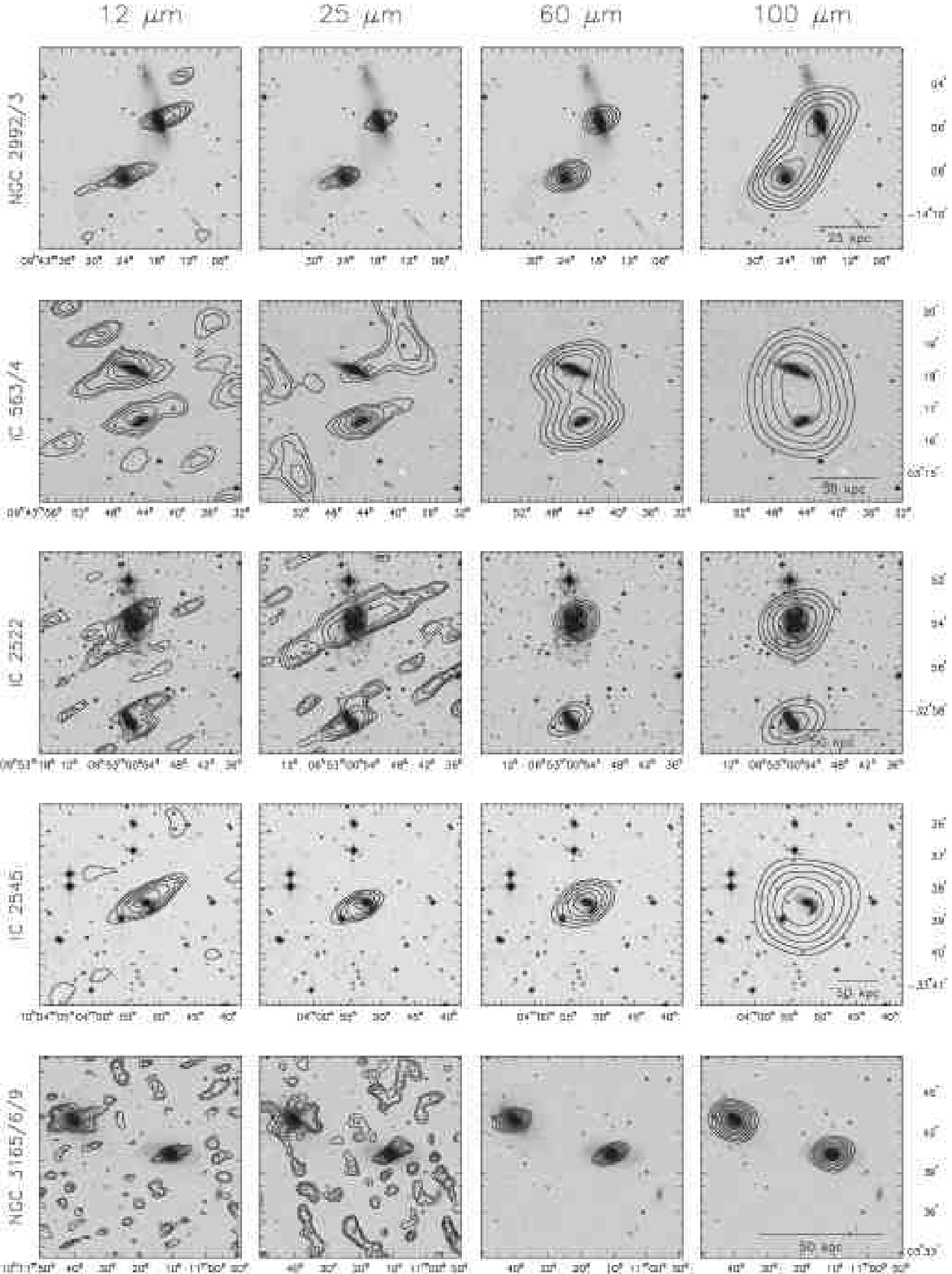}
            \caption{}
    \end{figure}
\clearpage  

\begin{figure}
    \epsscale{0.9}
    \figurenum{1 cont}
    \plotone{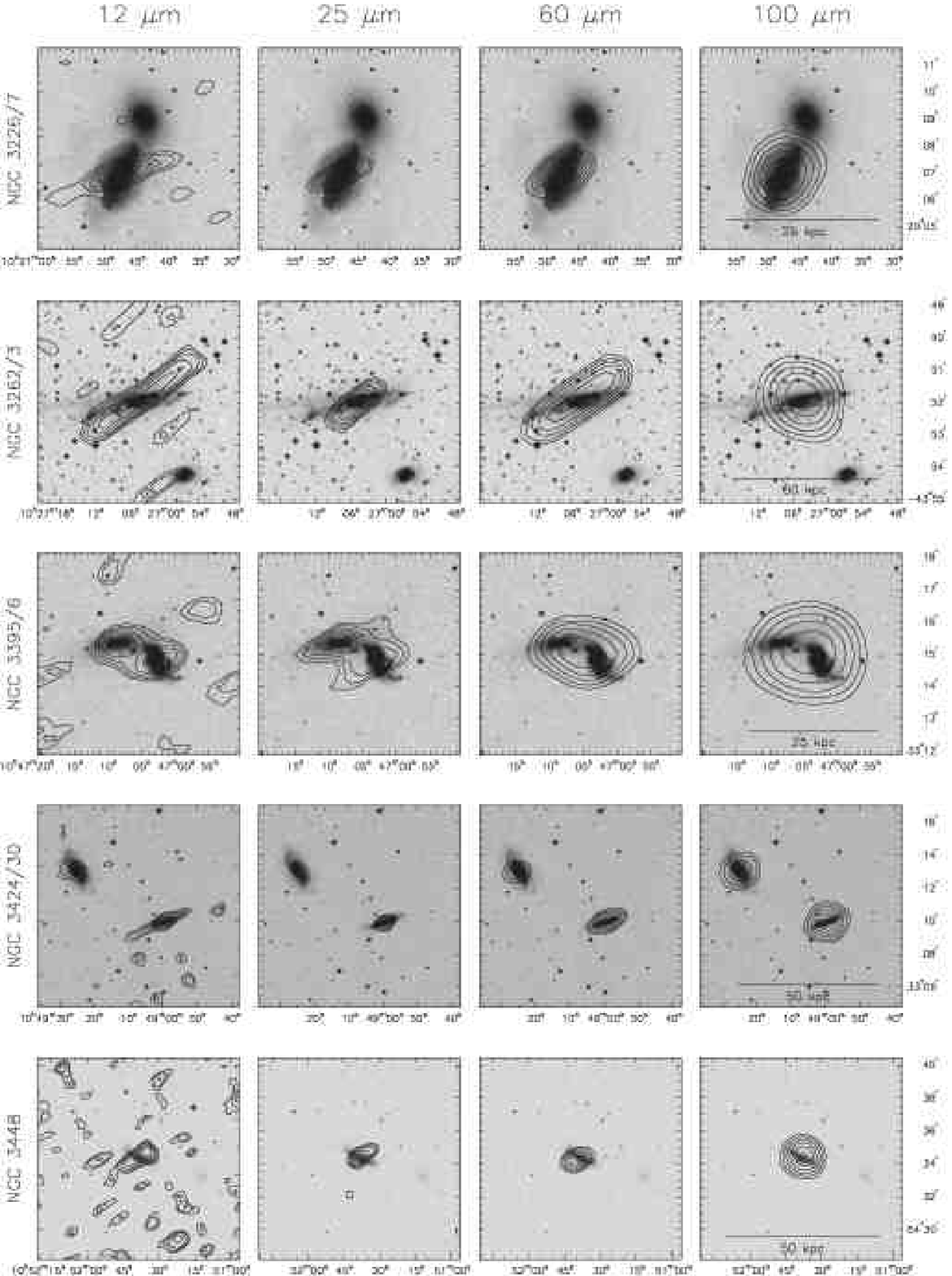}
            \caption{}
    \end{figure}
\clearpage  

\begin{figure}
    \epsscale{0.9}
    \figurenum{1 cont}
    \plotone{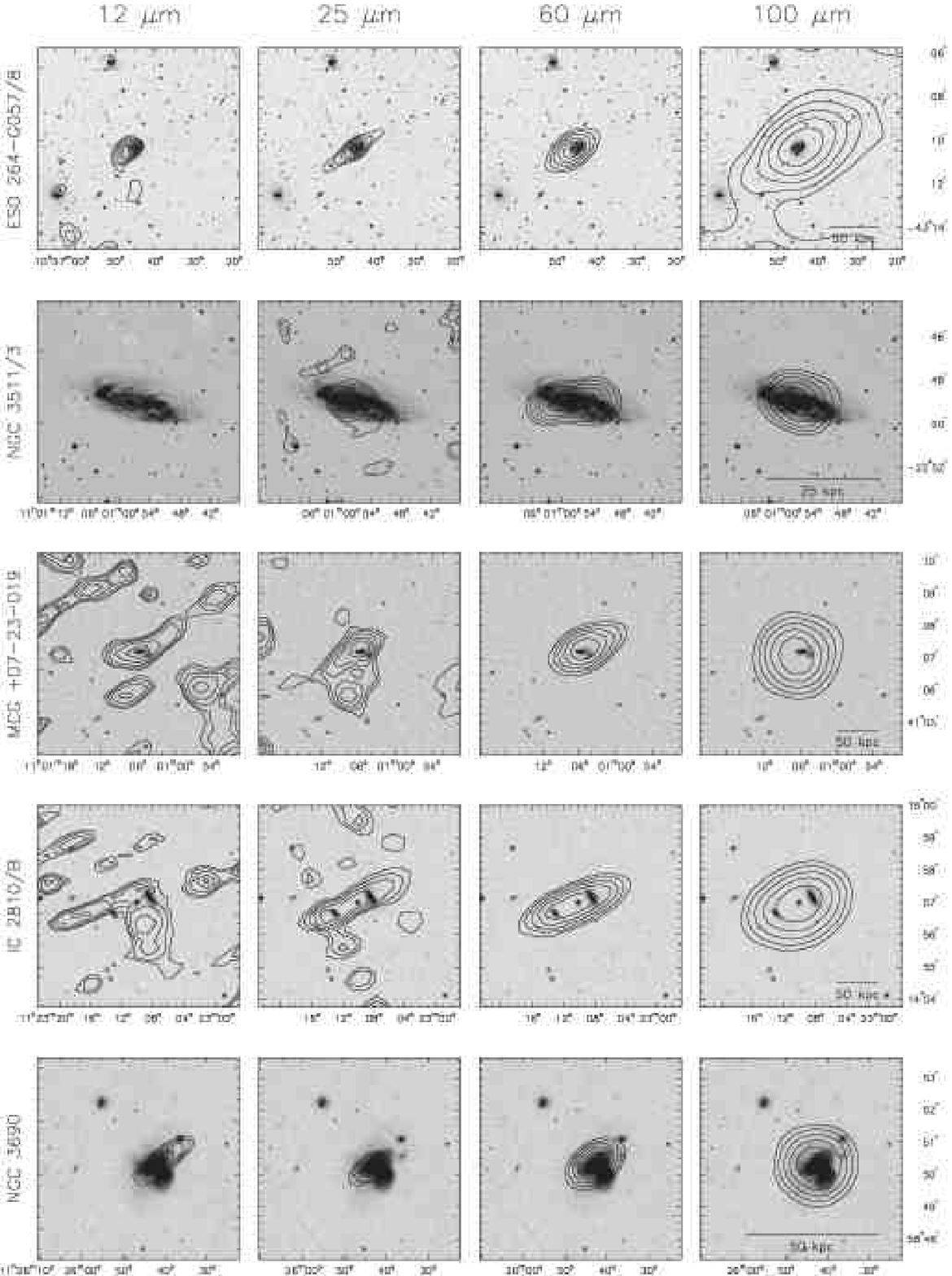}
            \caption{}
    \end{figure}
\clearpage  

\begin{figure}
    \epsscale{0.9}
    \figurenum{1 cont}
    \plotone{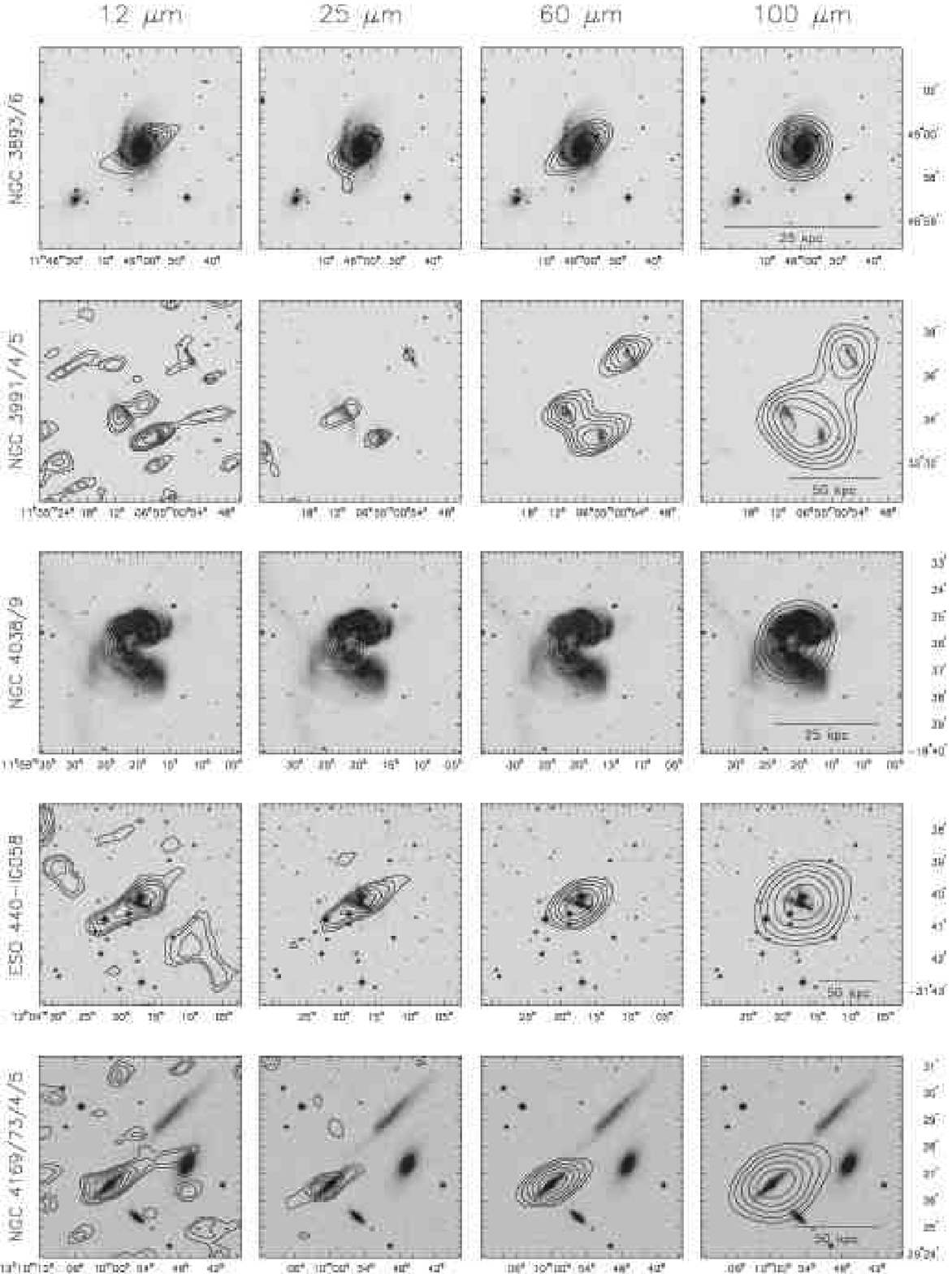}
            \caption{}
    \end{figure}
\clearpage  

\begin{figure}
    \epsscale{0.9}
    \figurenum{1 cont}
    \plotone{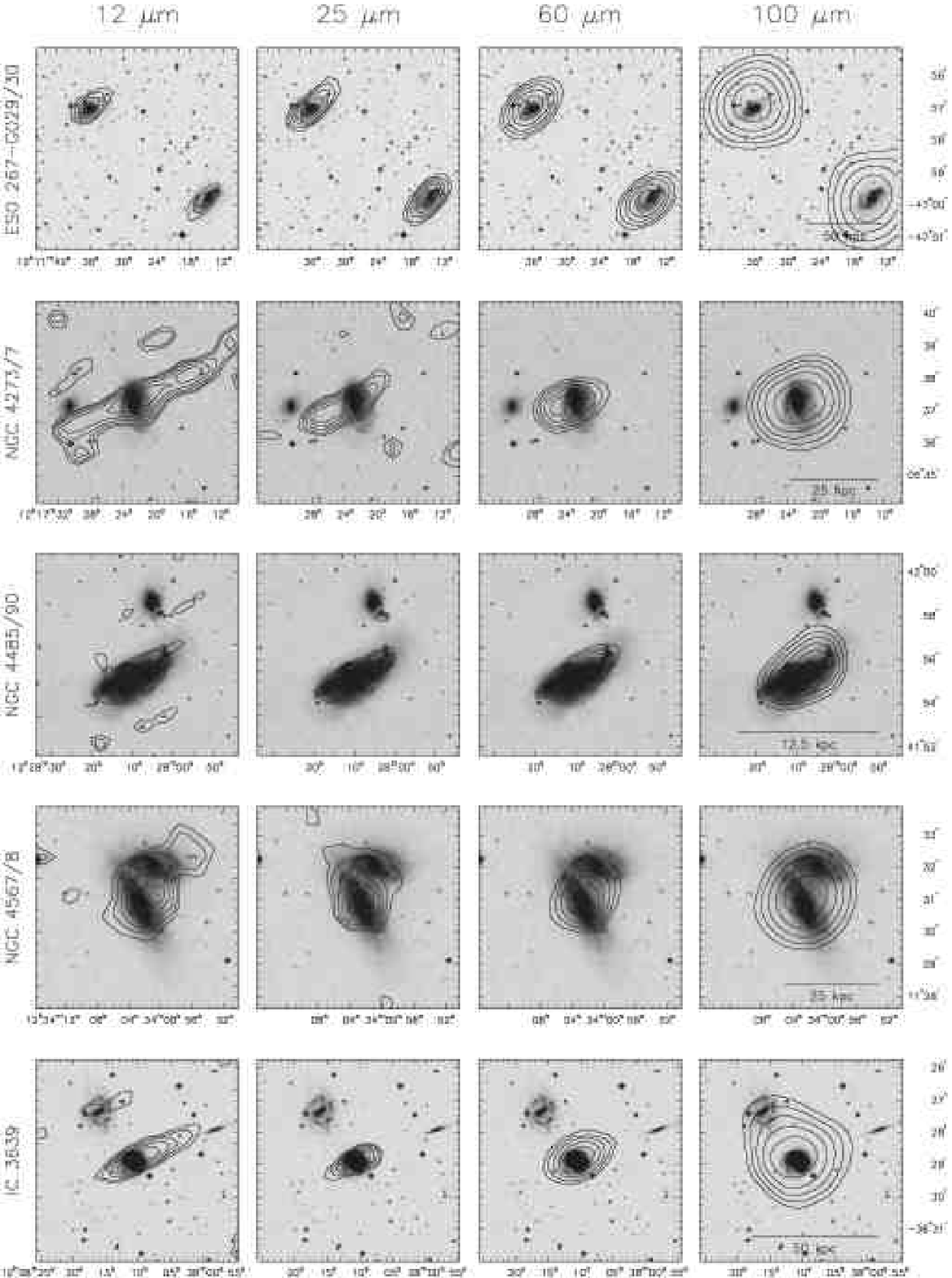}
            \caption{}
    \end{figure}
\clearpage  

\begin{figure}
    \epsscale{0.9}
    \figurenum{1 cont}
    \plotone{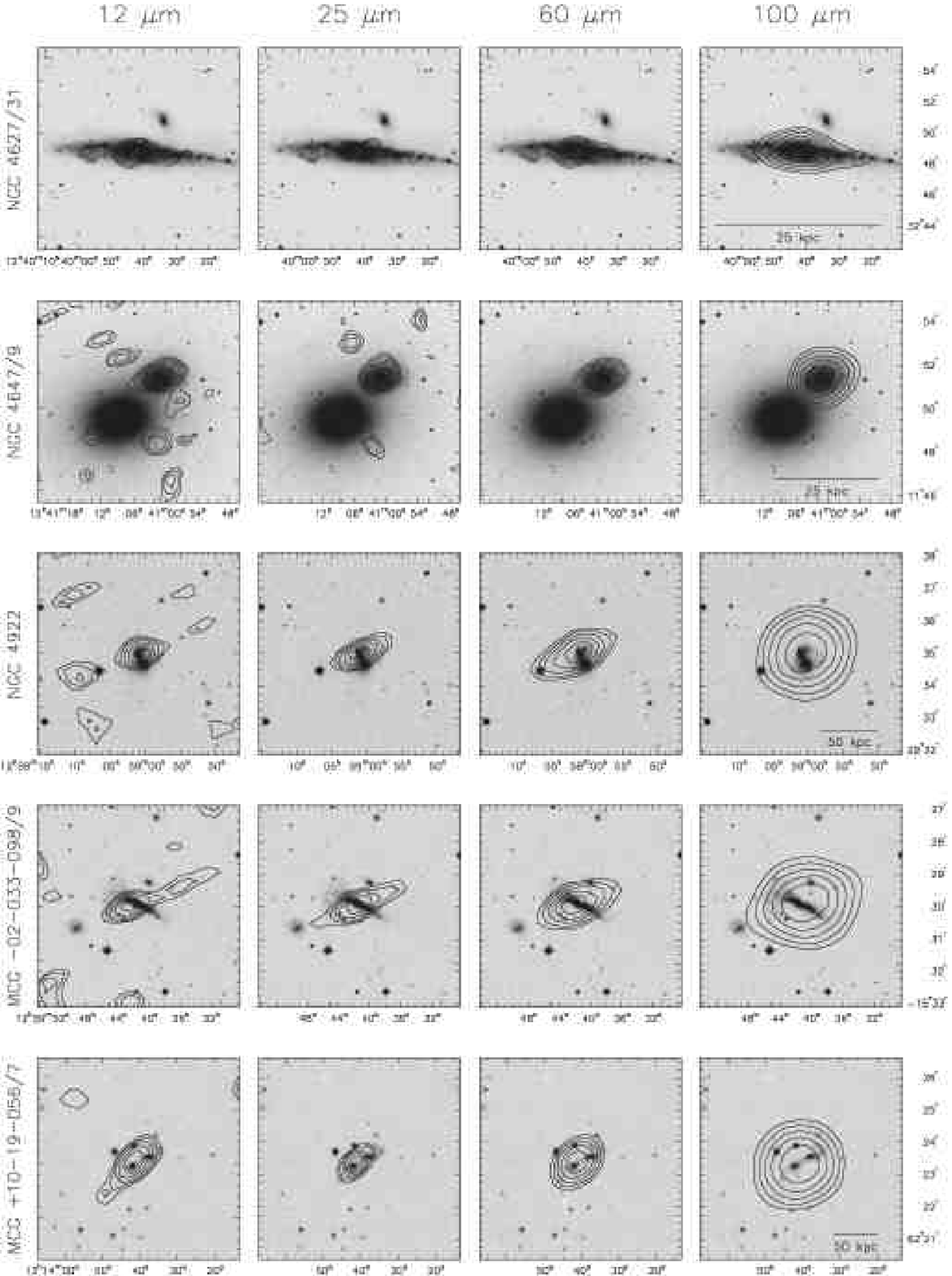}
            \caption{}
    \end{figure}
\clearpage  

\begin{figure}
    \epsscale{0.9}
    \figurenum{1 cont}
    \plotone{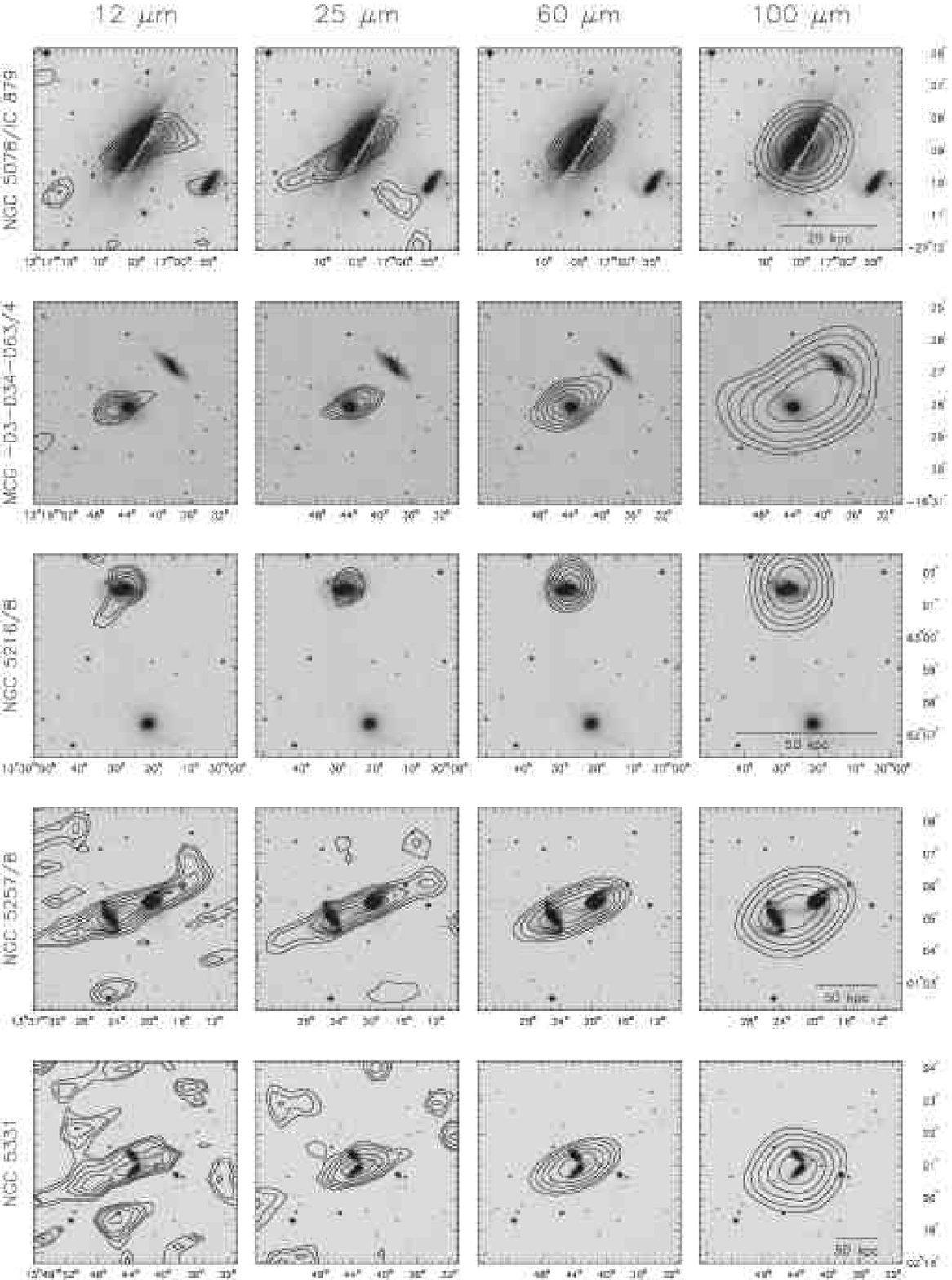}
            \caption{}
    \end{figure}
\clearpage  

\begin{figure}
    \epsscale{0.9}
    \figurenum{1 cont}
    \plotone{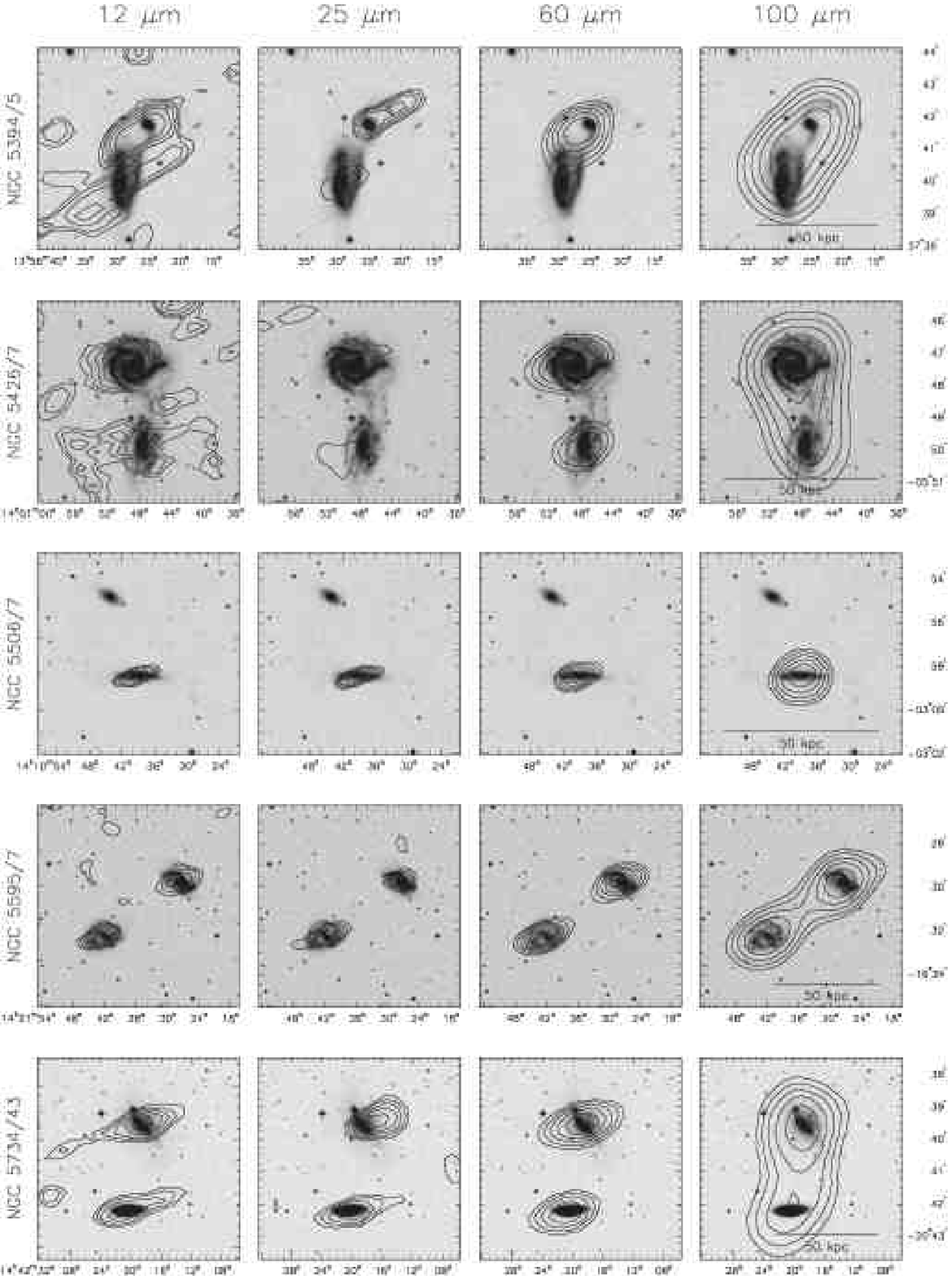}
            \caption{}
    \end{figure}
\clearpage  

\begin{figure}
    \epsscale{0.9}
    \figurenum{1 cont}
    \plotone{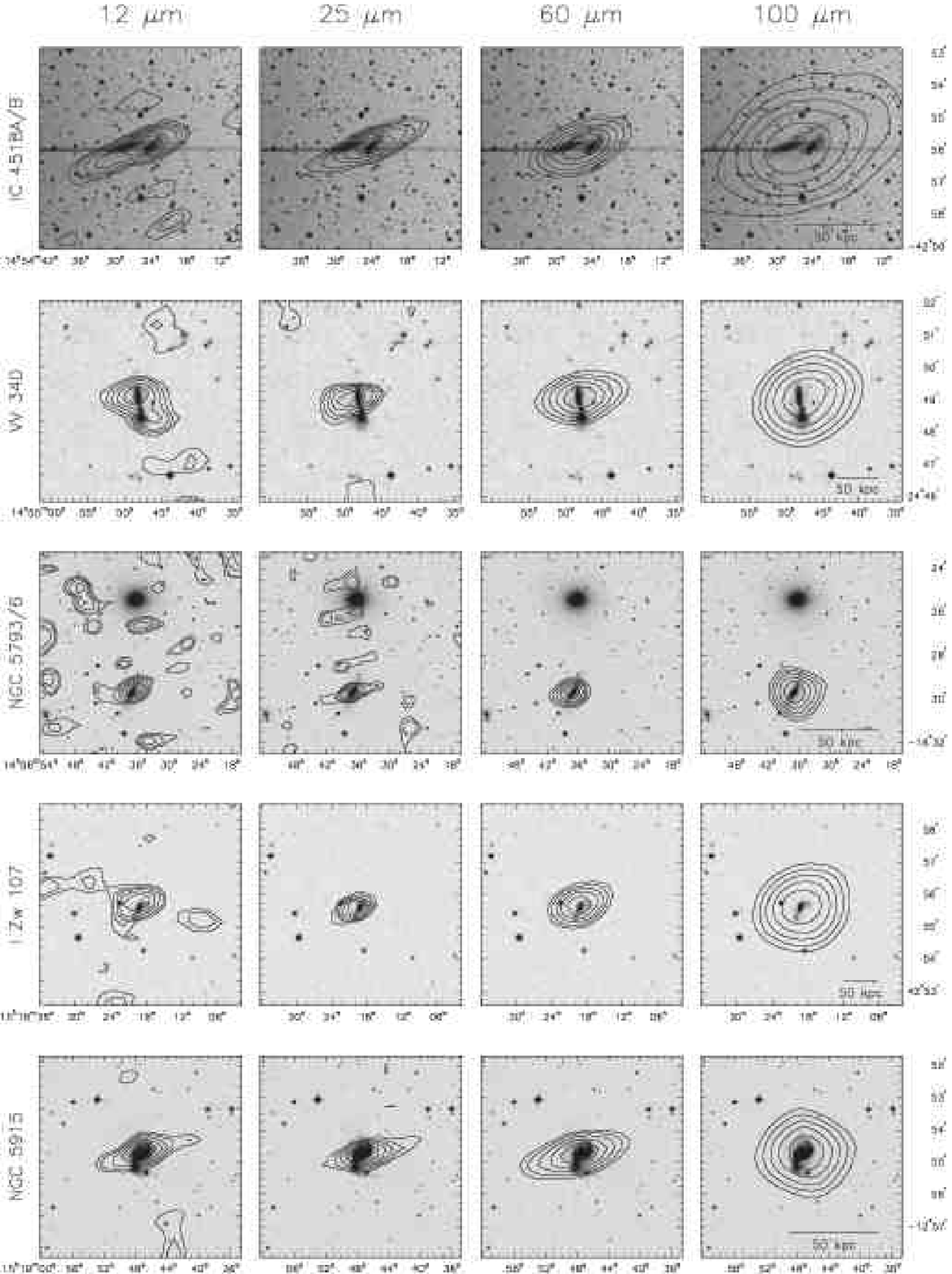}
            \caption{}
    \end{figure}
\clearpage  

\begin{figure}
    \epsscale{0.9}
    \figurenum{1 cont}
    \plotone{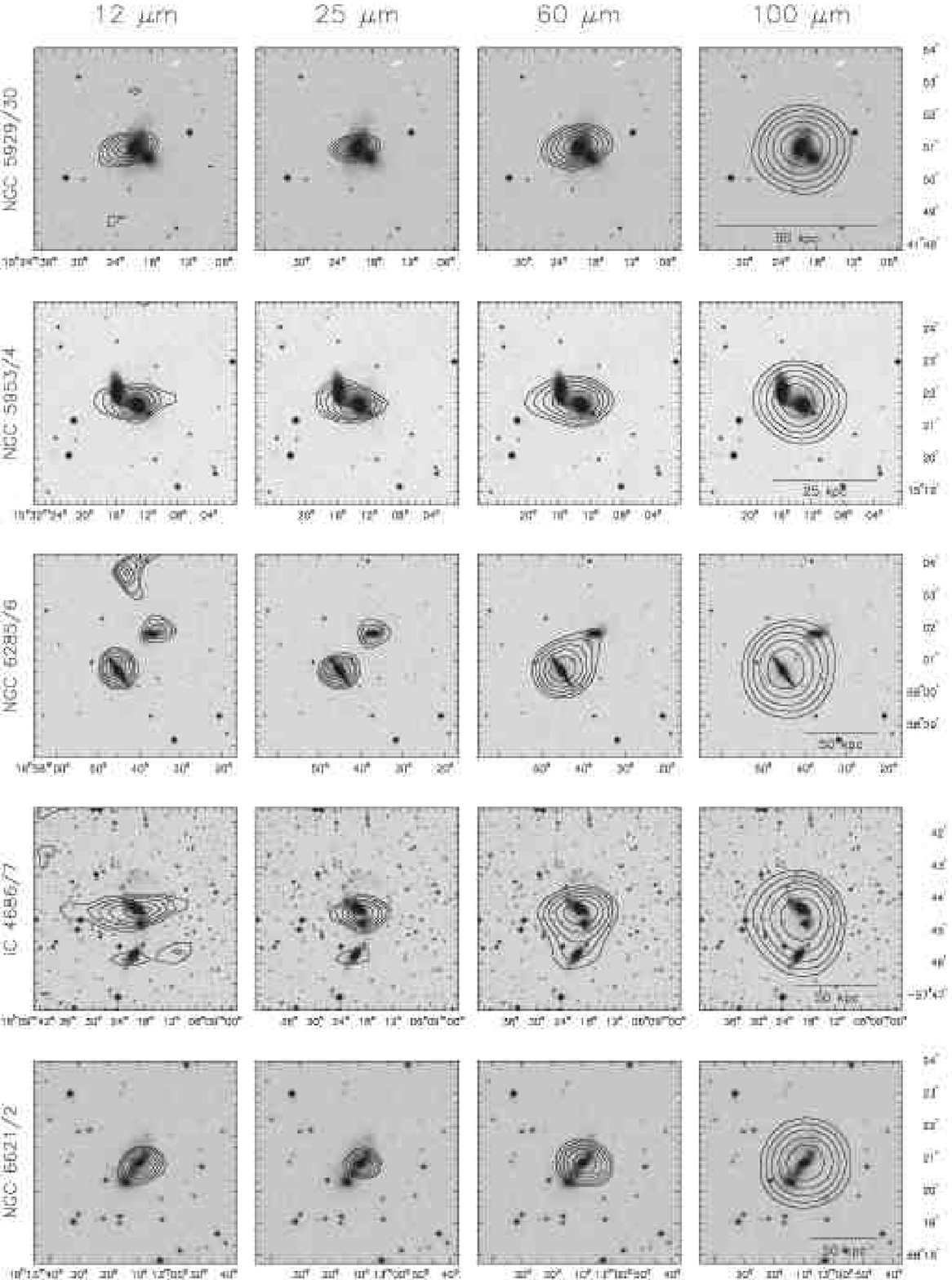}
            \caption{}
    \end{figure}
\clearpage  

\begin{figure}
    \epsscale{0.9}
    \figurenum{1 cont}
    \plotone{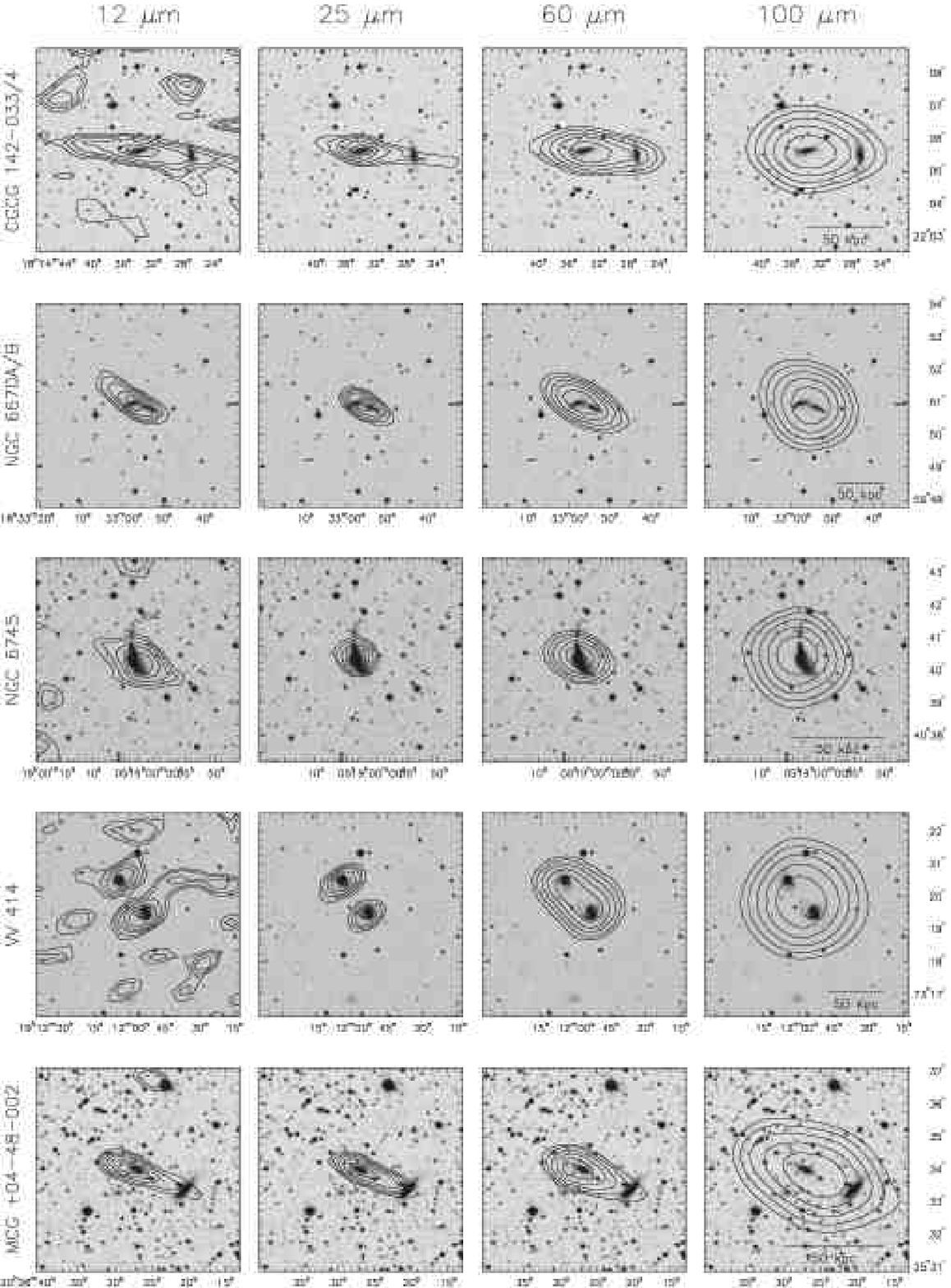}
            \caption{}
    \end{figure}
\clearpage  

\begin{figure}
    \epsscale{0.9}
    \figurenum{1 cont}
    \plotone{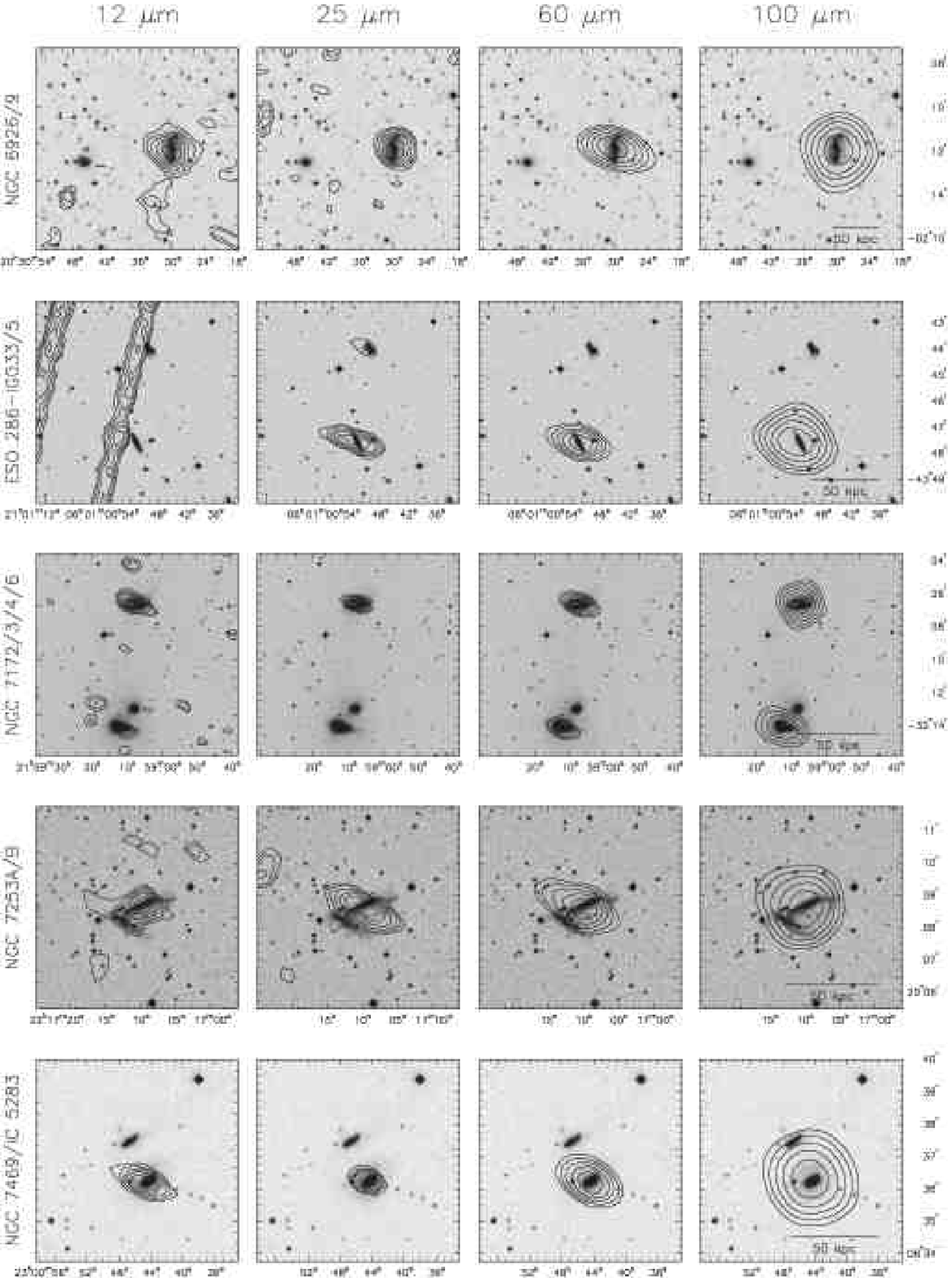}
            \caption{}
    \end{figure}
\clearpage  

\begin{figure}
    \epsscale{0.9}
    \figurenum{1 cont}
    \plotone{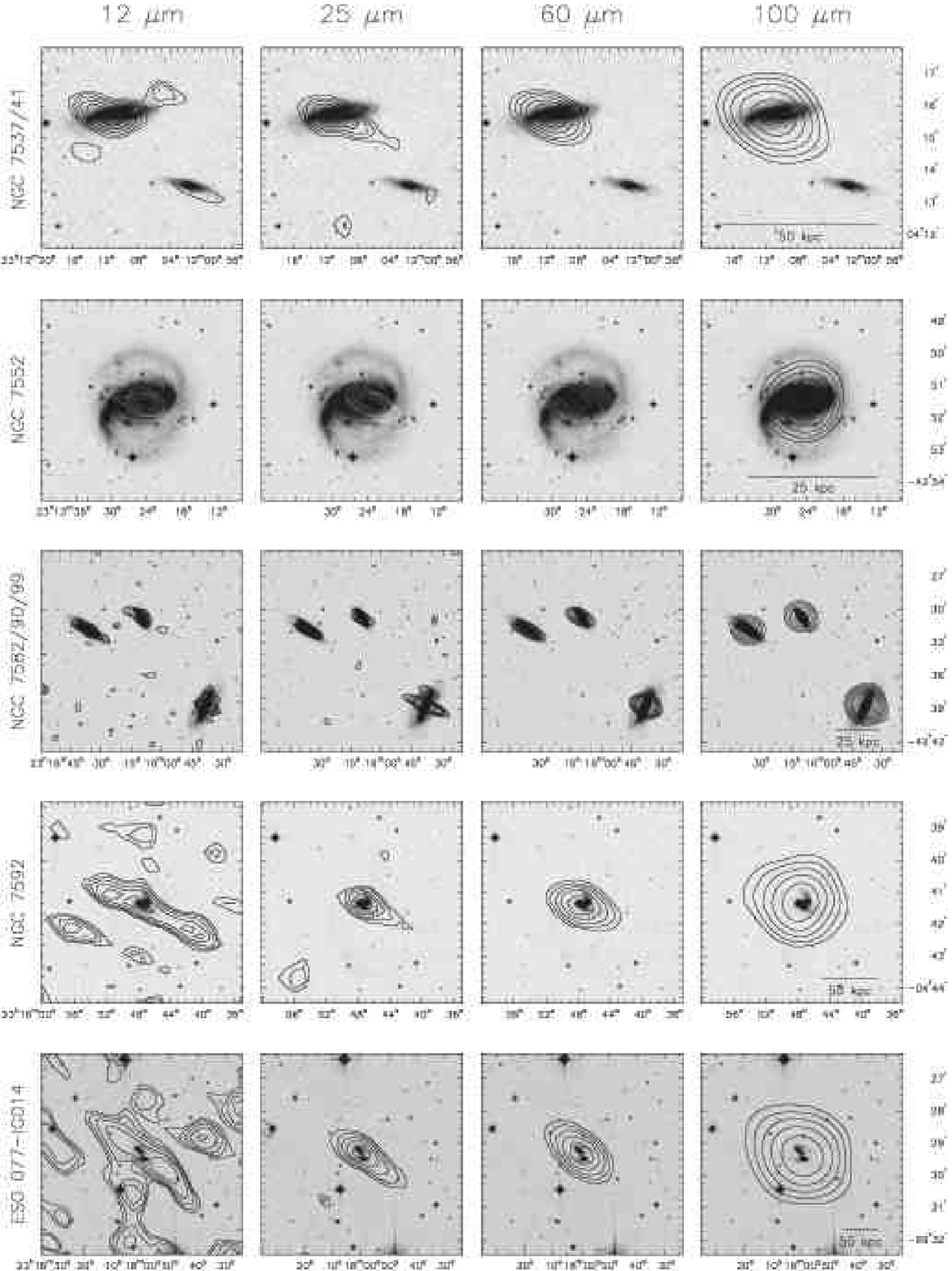}
            \caption{}
    \end{figure}
\clearpage  

\begin{figure}
    \epsscale{0.9}
    \figurenum{1 cont}
    \plotone{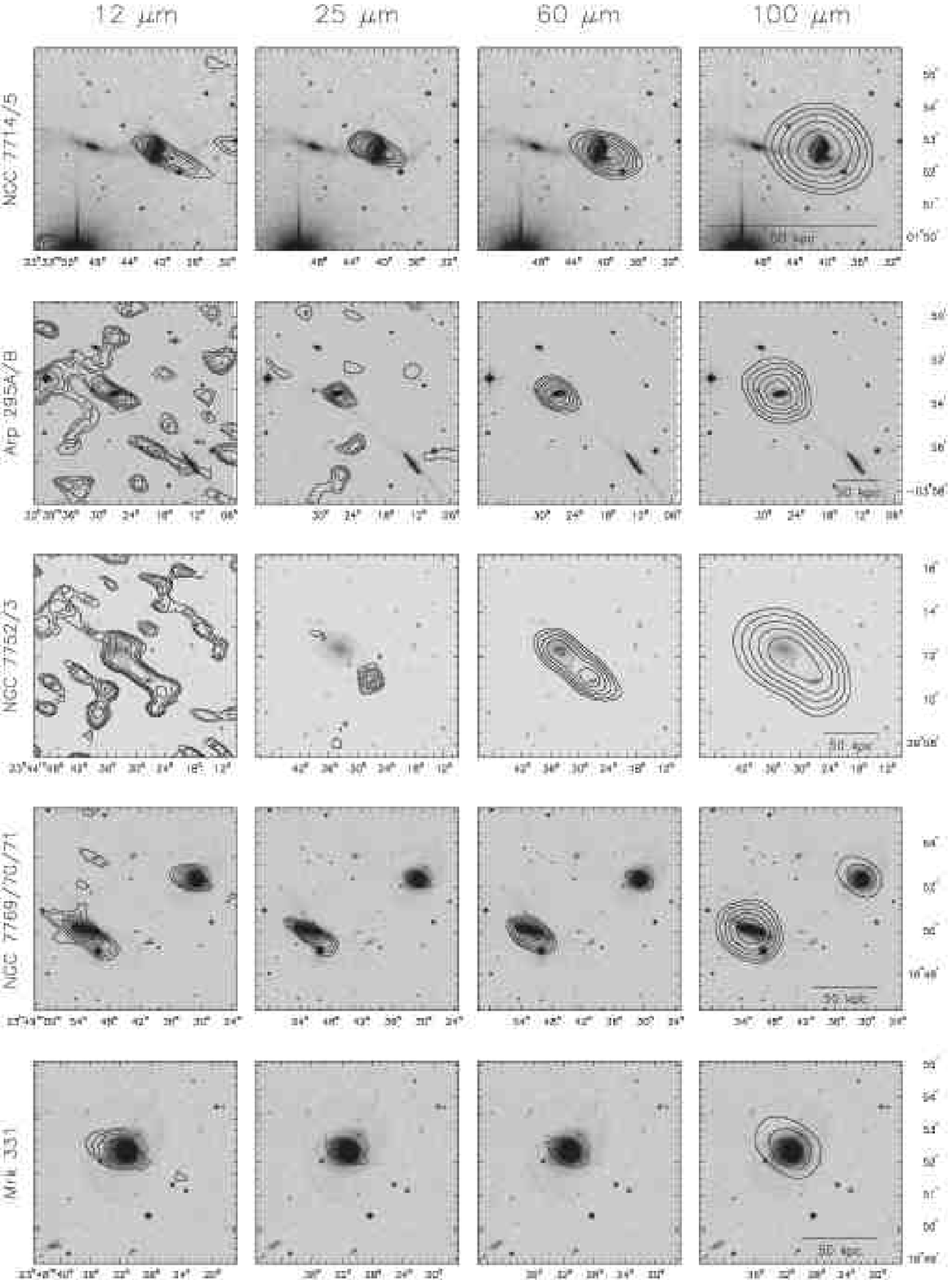}
            \caption{}
    \end{figure}
\clearpage  

\begin{figure}
    \epsscale{0.9}
    \figurenum{1 cont}
    \plotone{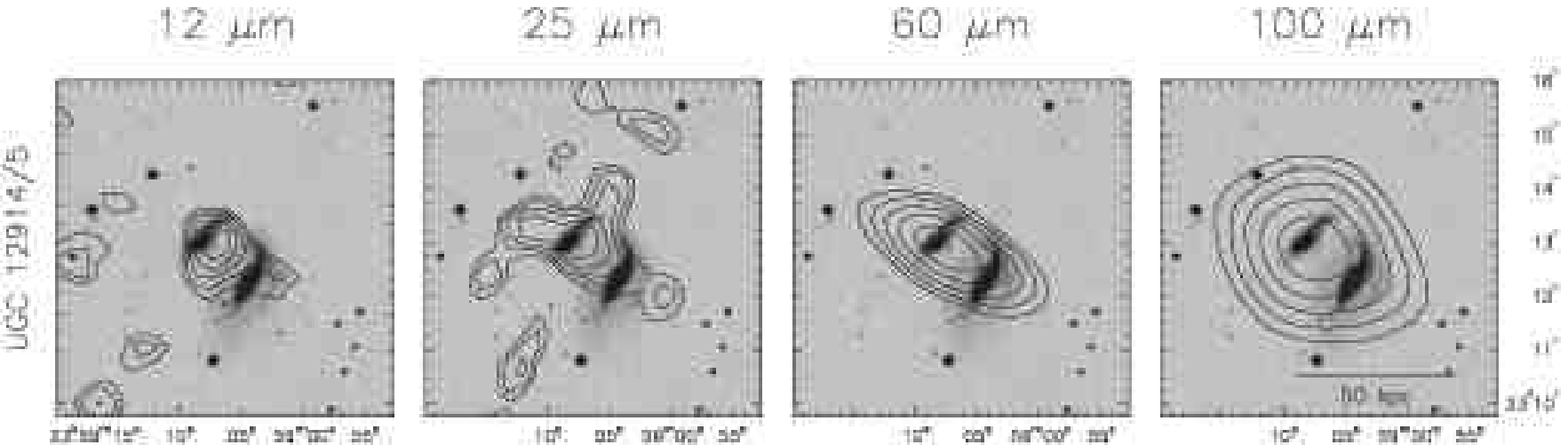}
            \caption{}
    \end{figure}
\clearpage  

\setcounter{figure}{1}
\begin{figure}
    \epsscale{0.5}
    \plotone{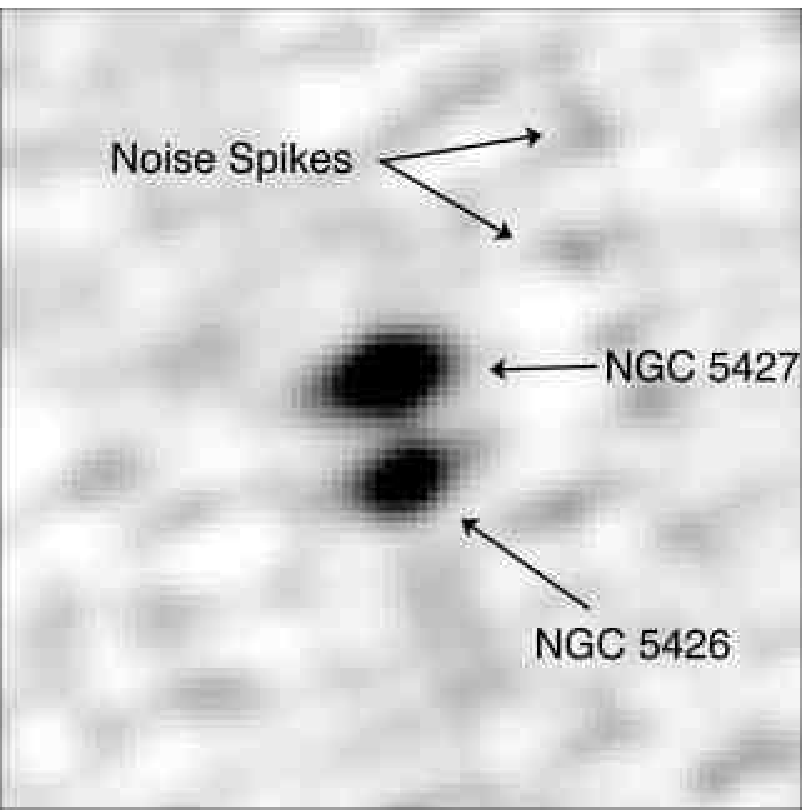}
        \caption{}
    \end{figure}
\clearpage

\begin{figure}
    \epsscale{0.7}
    \plotone{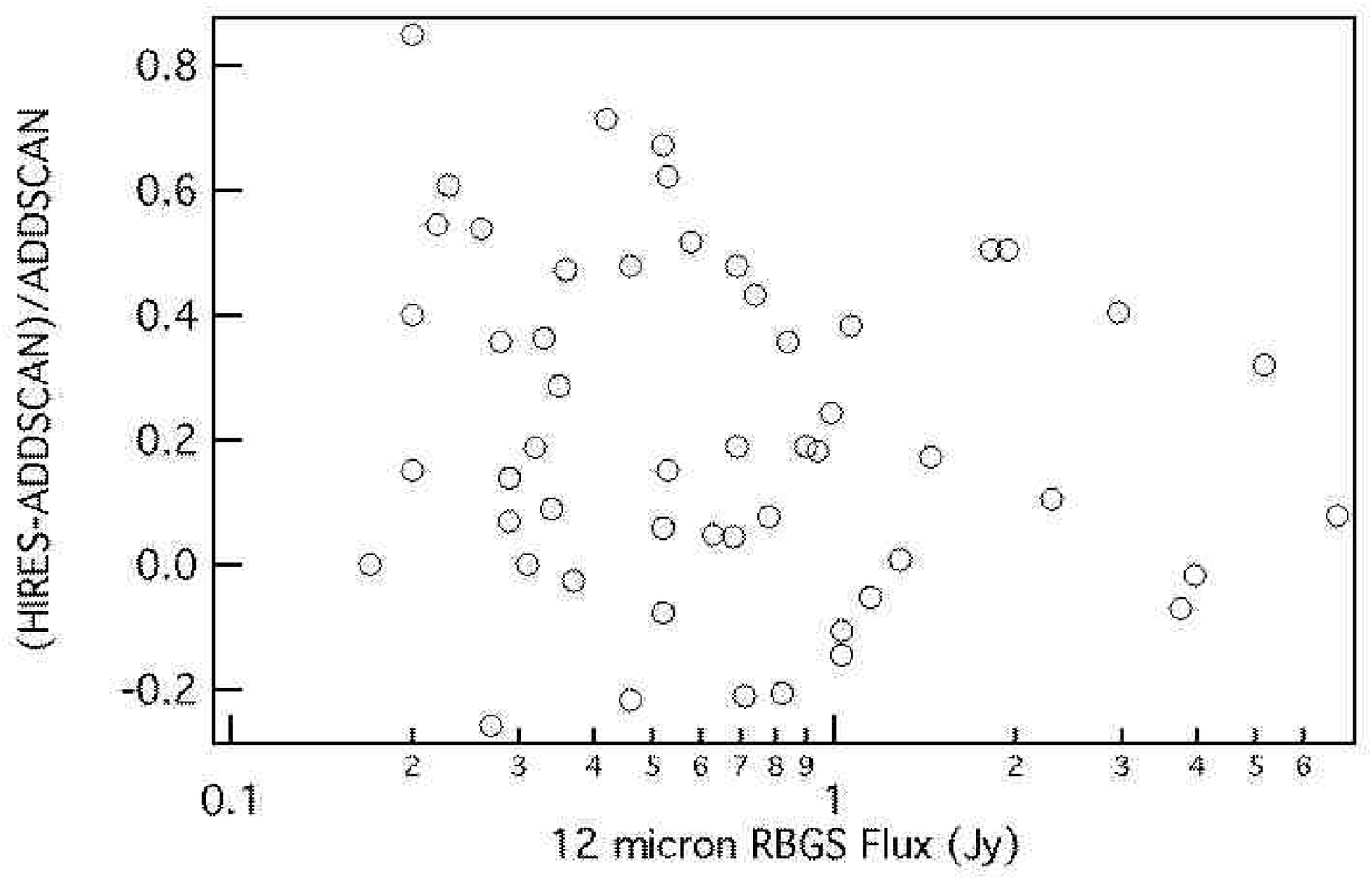}
        \caption{}
    \end{figure}
%\clearpage  

\begin{figure}
    \epsscale{0.7}
    \plotone{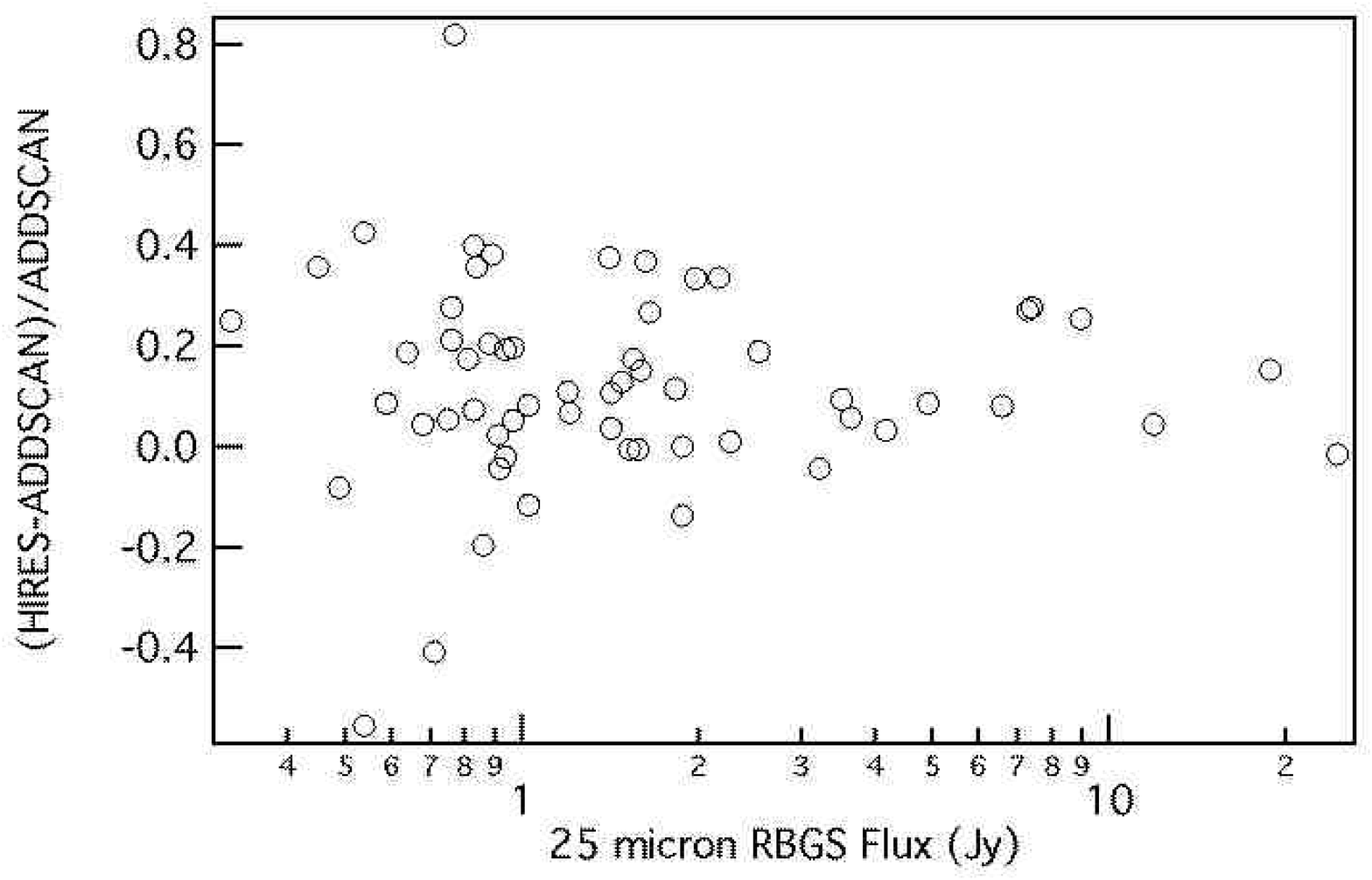}
        \caption{}
    \end{figure}
\clearpage  
\begin{figure}
    \epsscale{0.7}
    \plotone{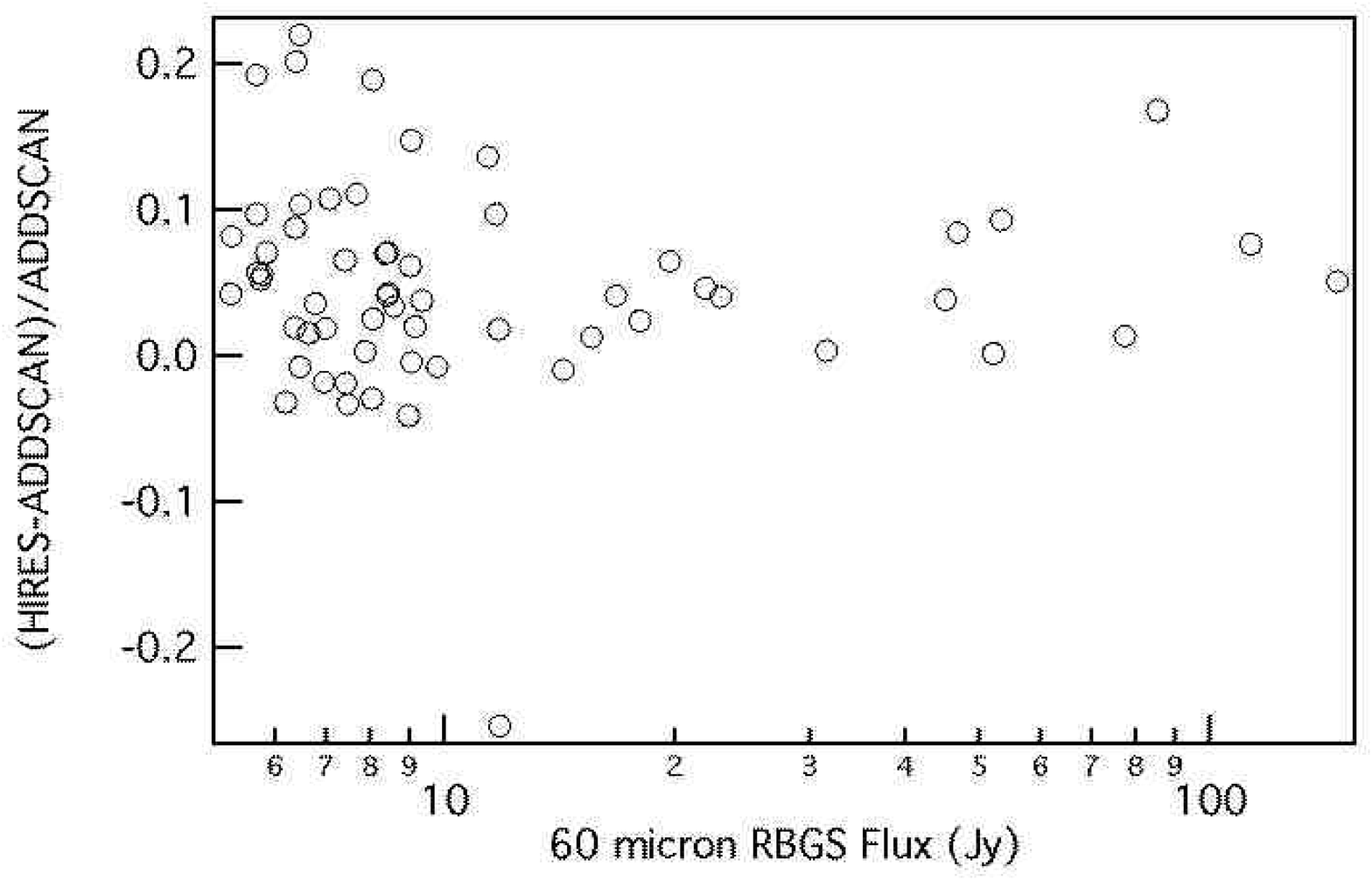}
        \caption{}
    \end{figure}
%\clearpage  
\begin{figure}
    \epsscale{0.7}
    \plotone{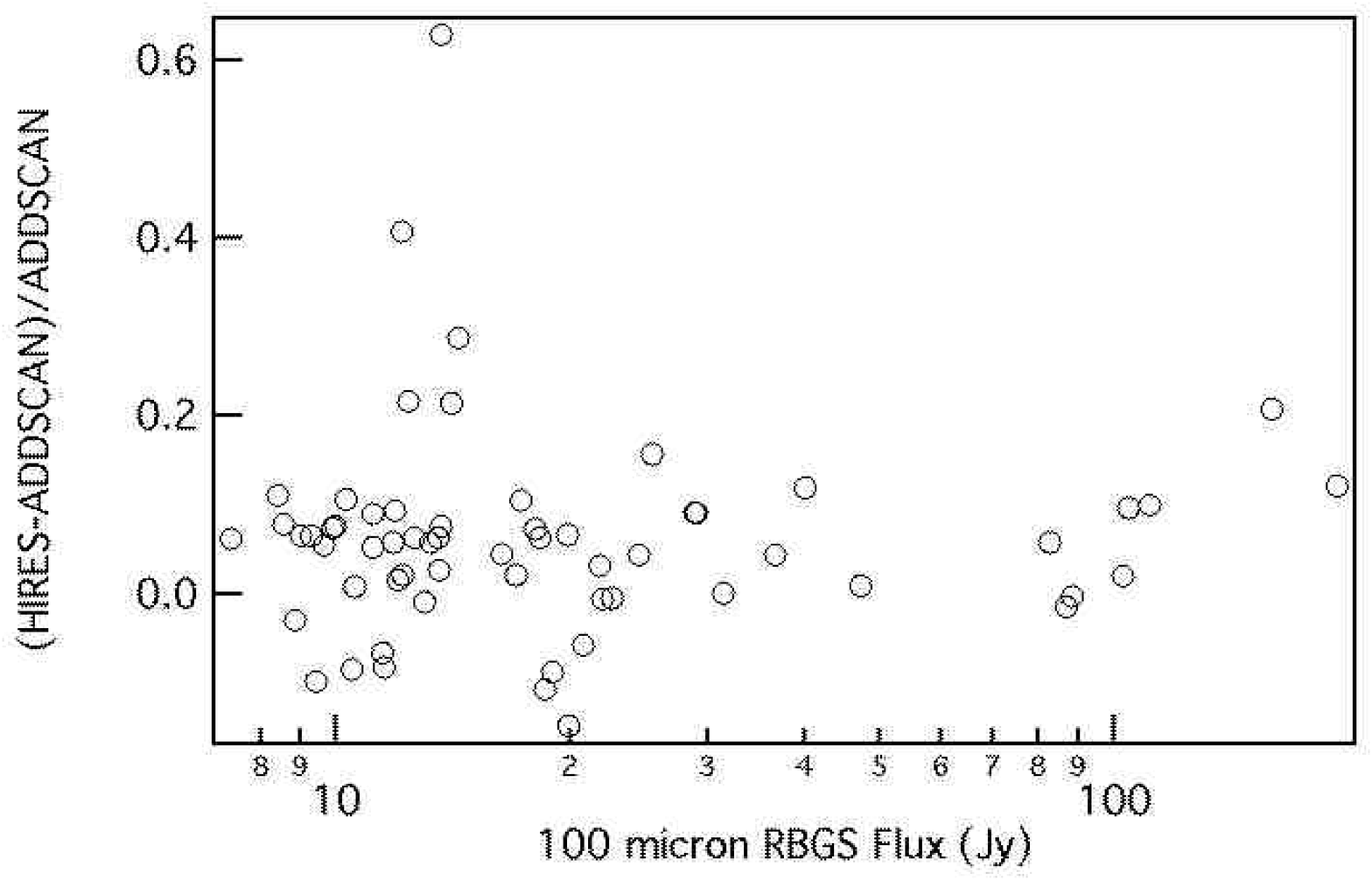}
        \caption{}
    \end{figure}
\clearpage  
\begin{figure}
    \epsscale{0.4}
    \plotone{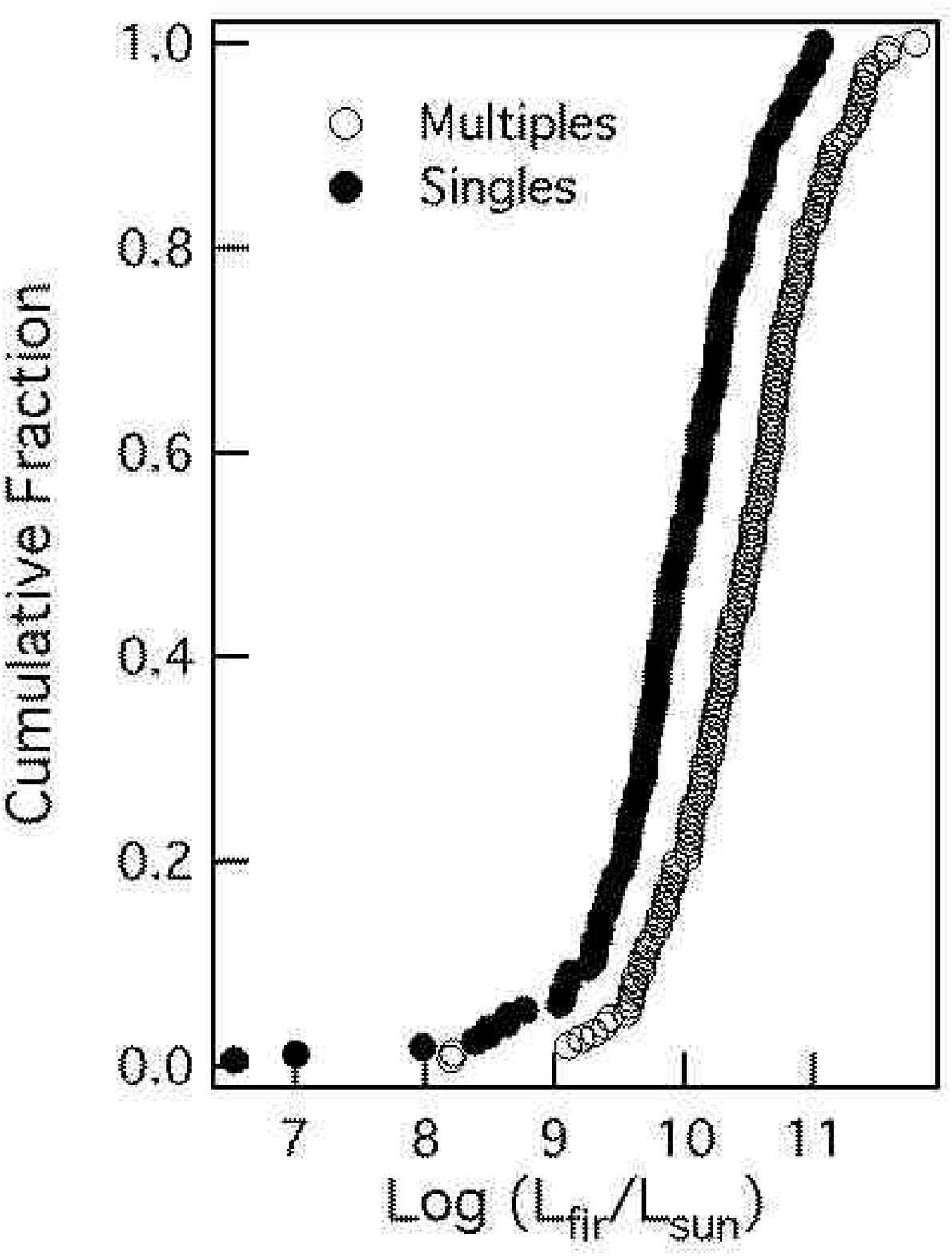}
        \caption{}
    \end{figure}
%\clearpage  
\begin{figure}
    \epsscale{0.4}
    \plotone{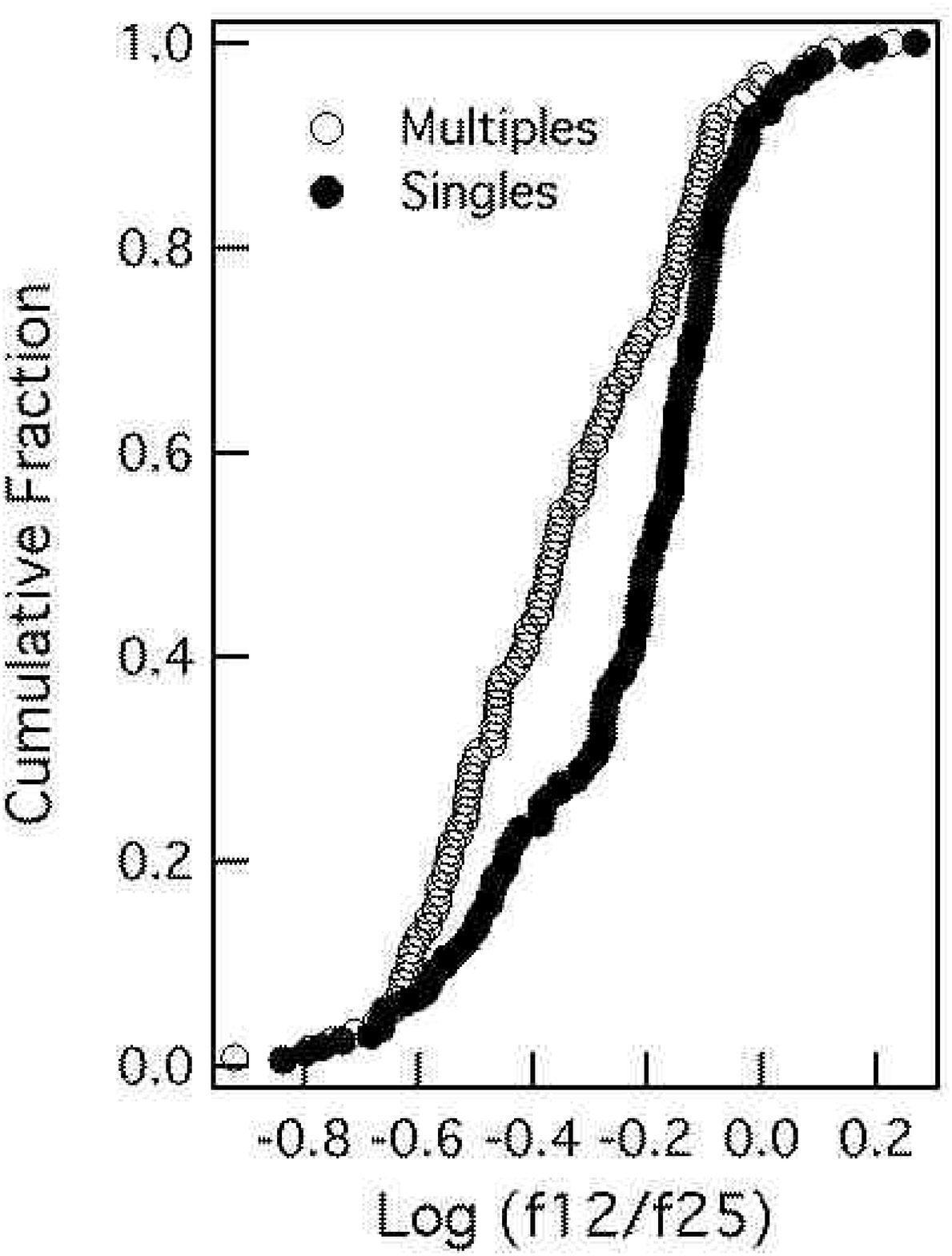}
        \caption{}
    \end{figure}
\clearpage  
\begin{figure}
    \epsscale{0.4}
    \plotone{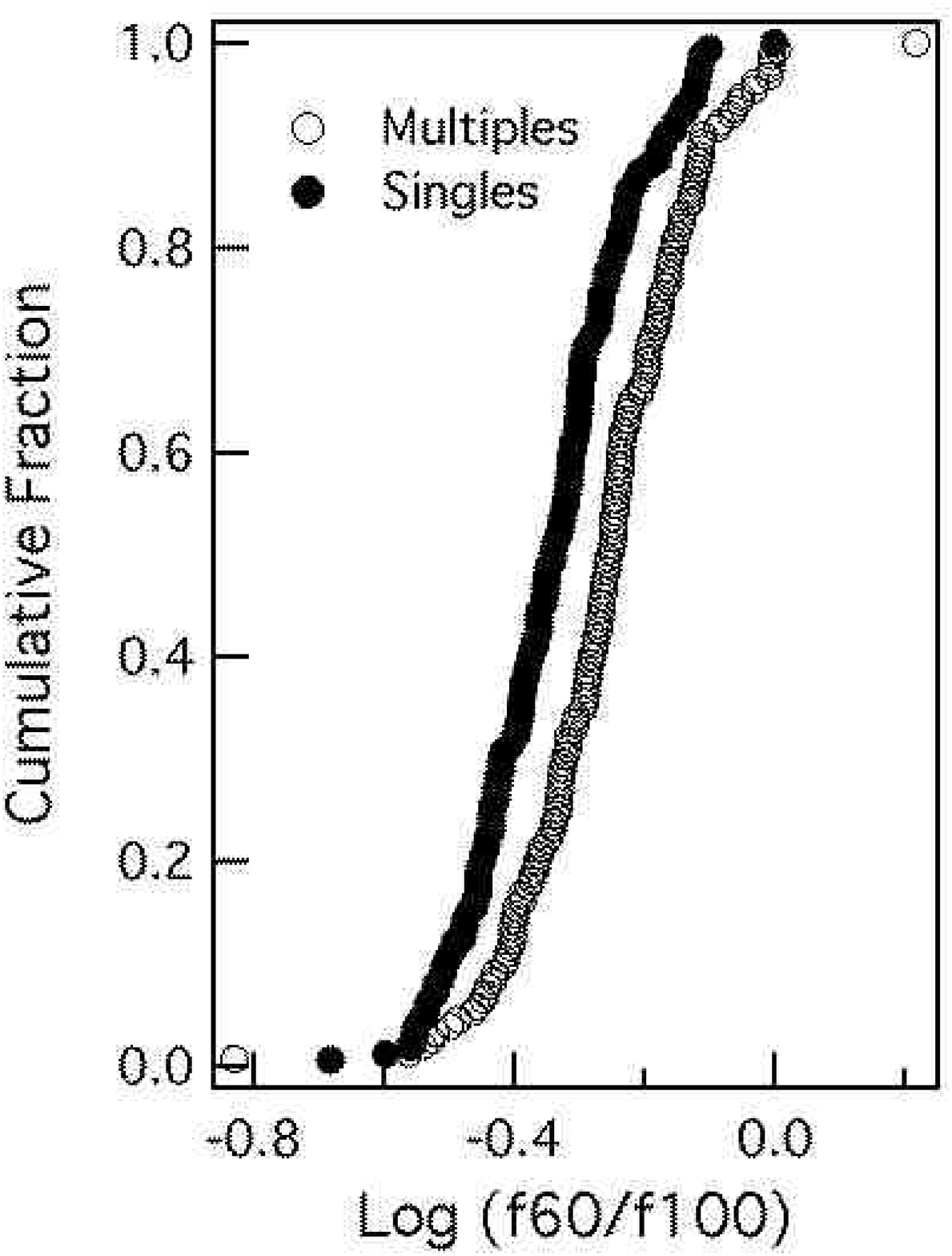}
        \caption{}
    \end{figure}
%\clearpage  
\begin{figure}
    \epsscale{0.5}
    \plotone{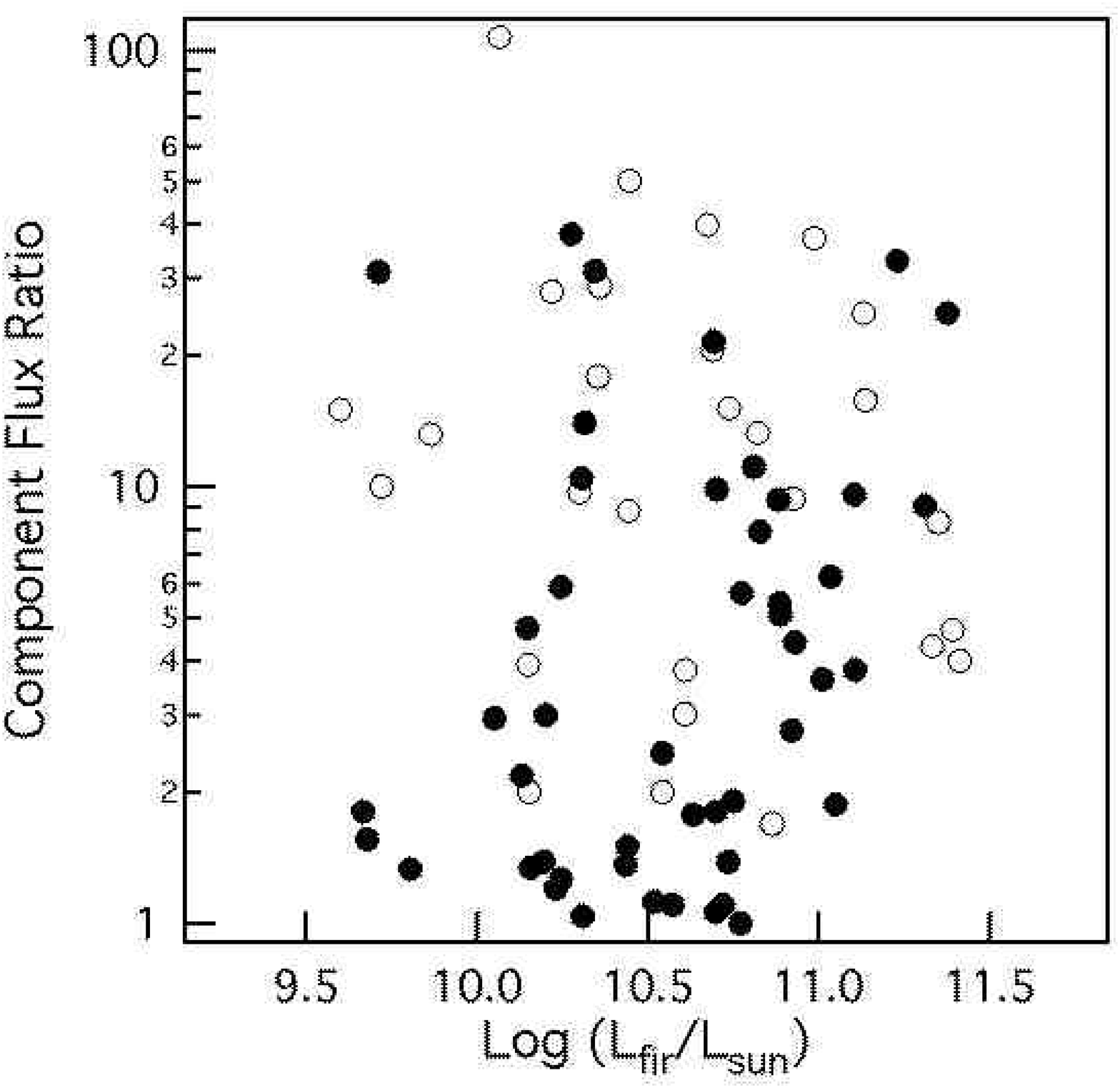}
        \caption{}
    \end{figure}
\clearpage  
\begin{figure}
    \epsscale{0.4}
    \plotone{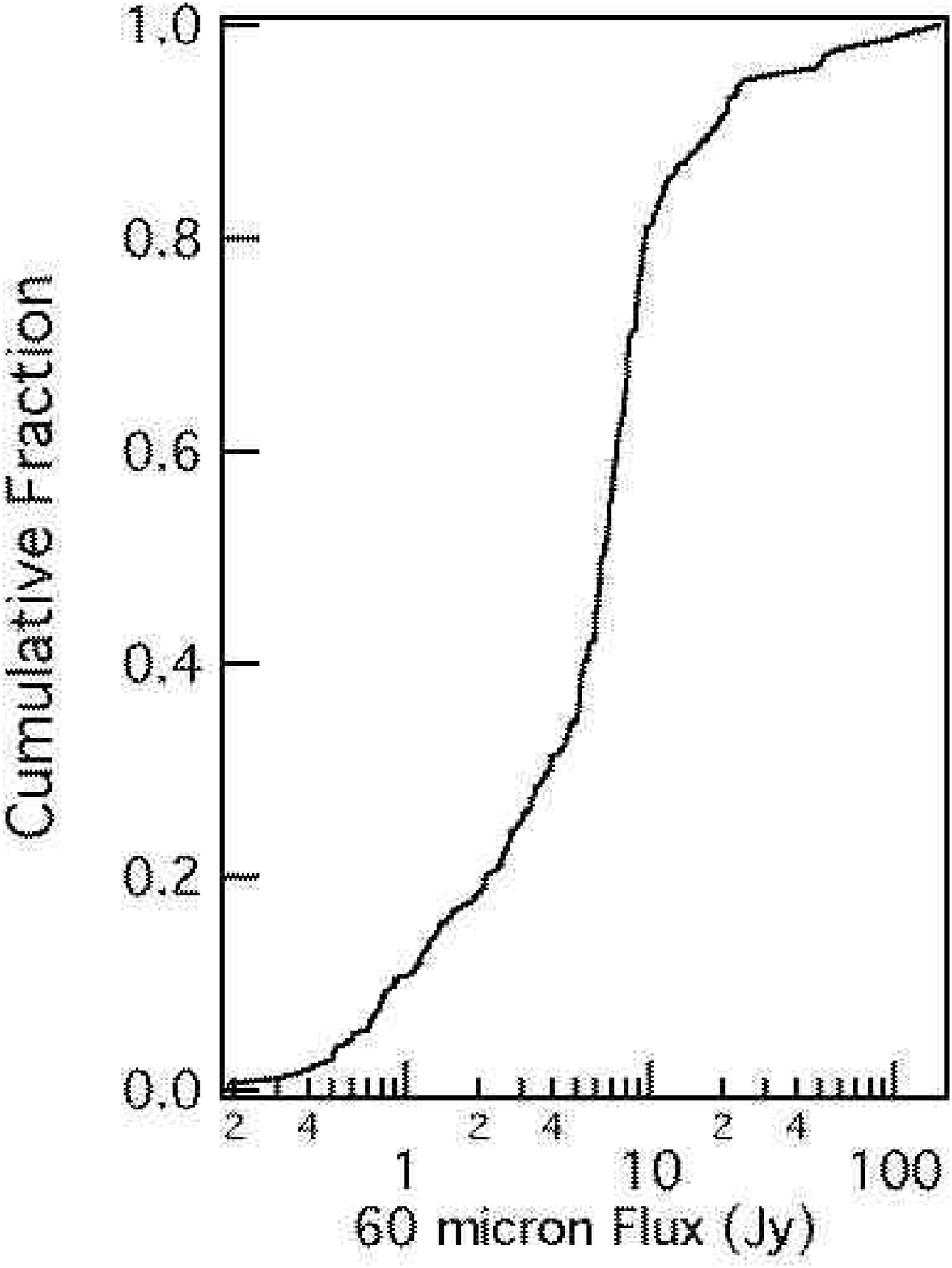}
        \caption{}
    \end{figure}
%\clearpage  
\begin{figure}
    \epsscale{0.4}
    \plotone{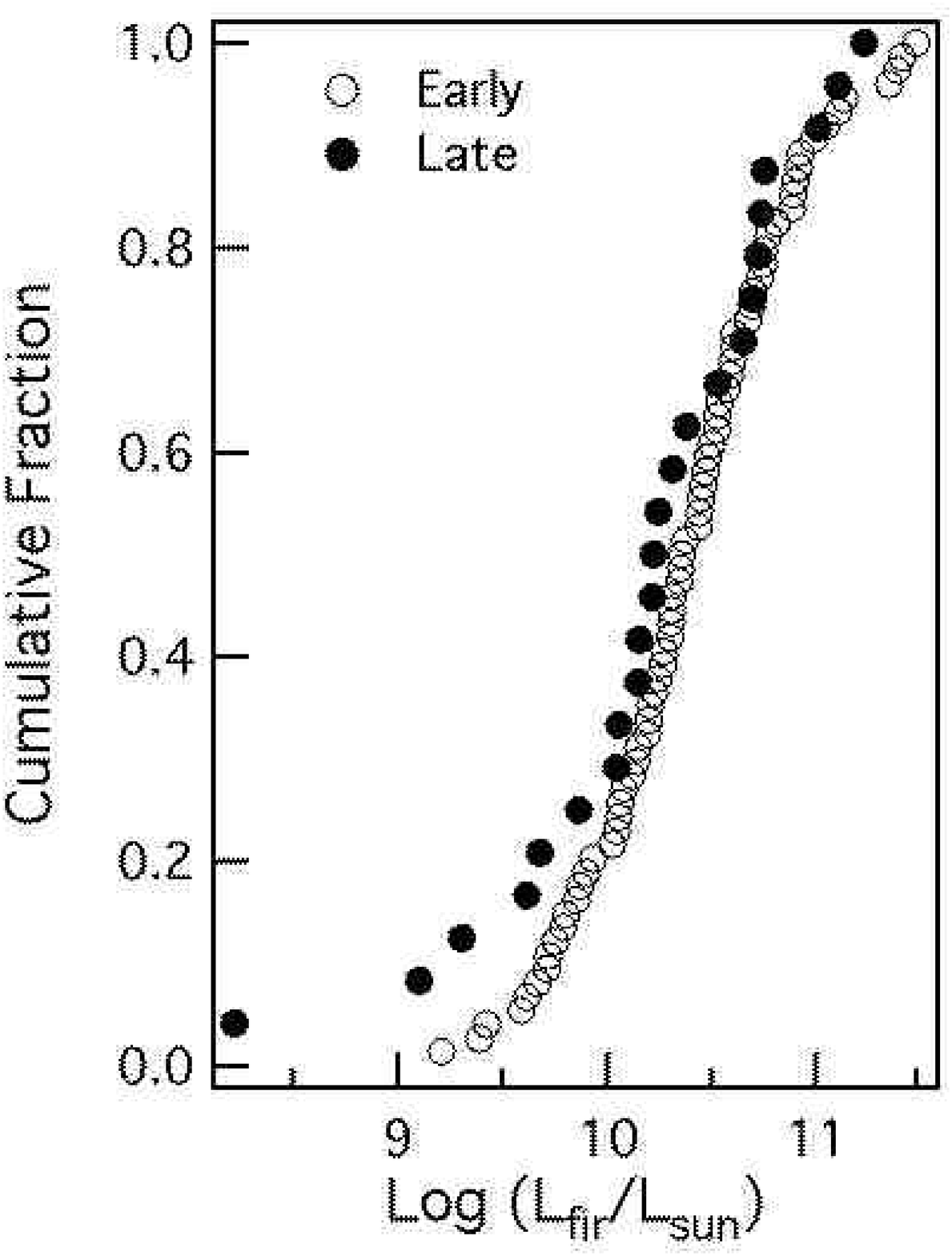}
        \caption{}
    \end{figure}
\clearpage  

\begin{figure}
    \epsscale{0.9}
    \plotone{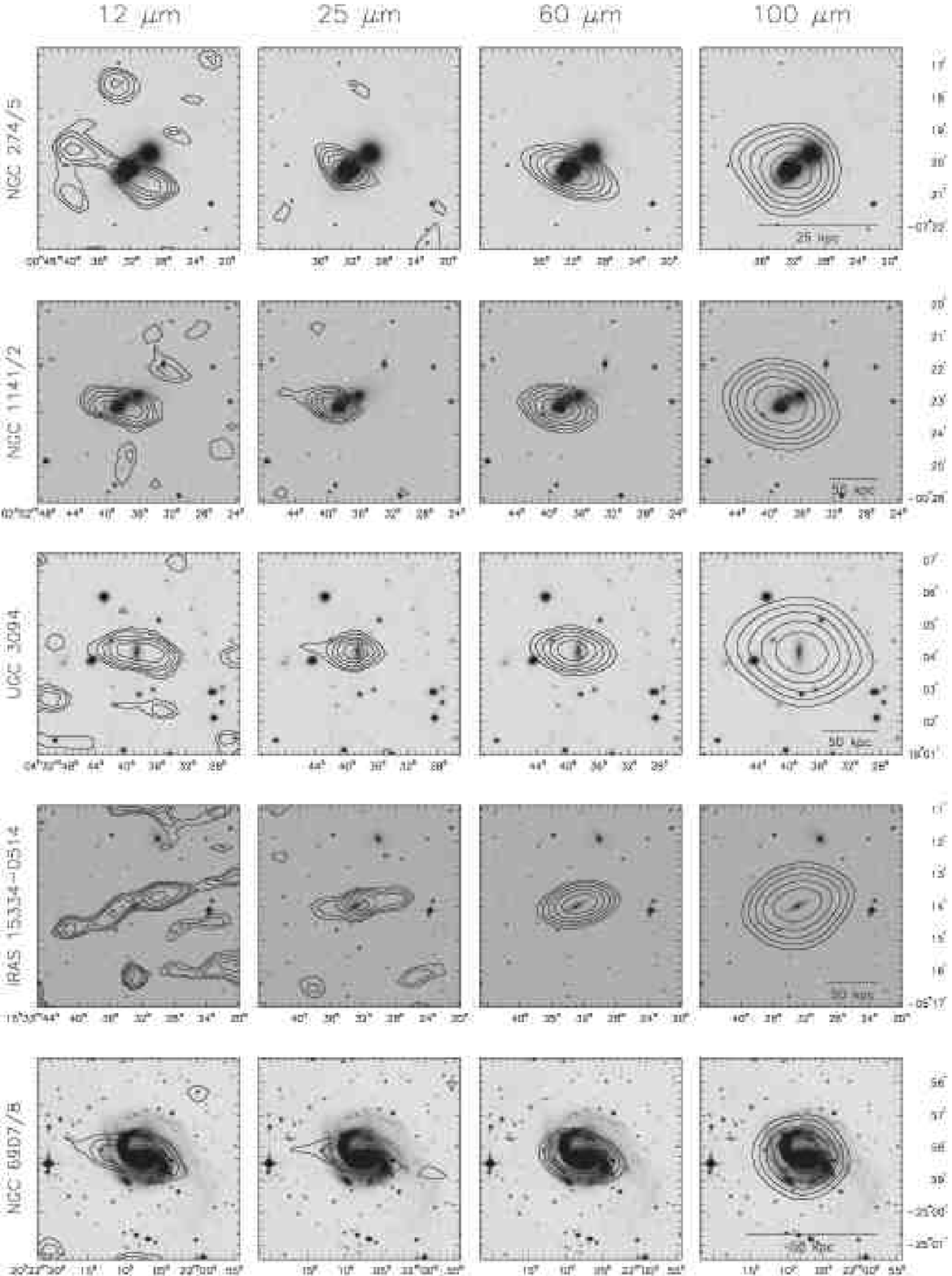}
            \caption{}
    \end{figure}

\clearpage  

\begin{figure}
    \epsscale{1.1}
     \plotone{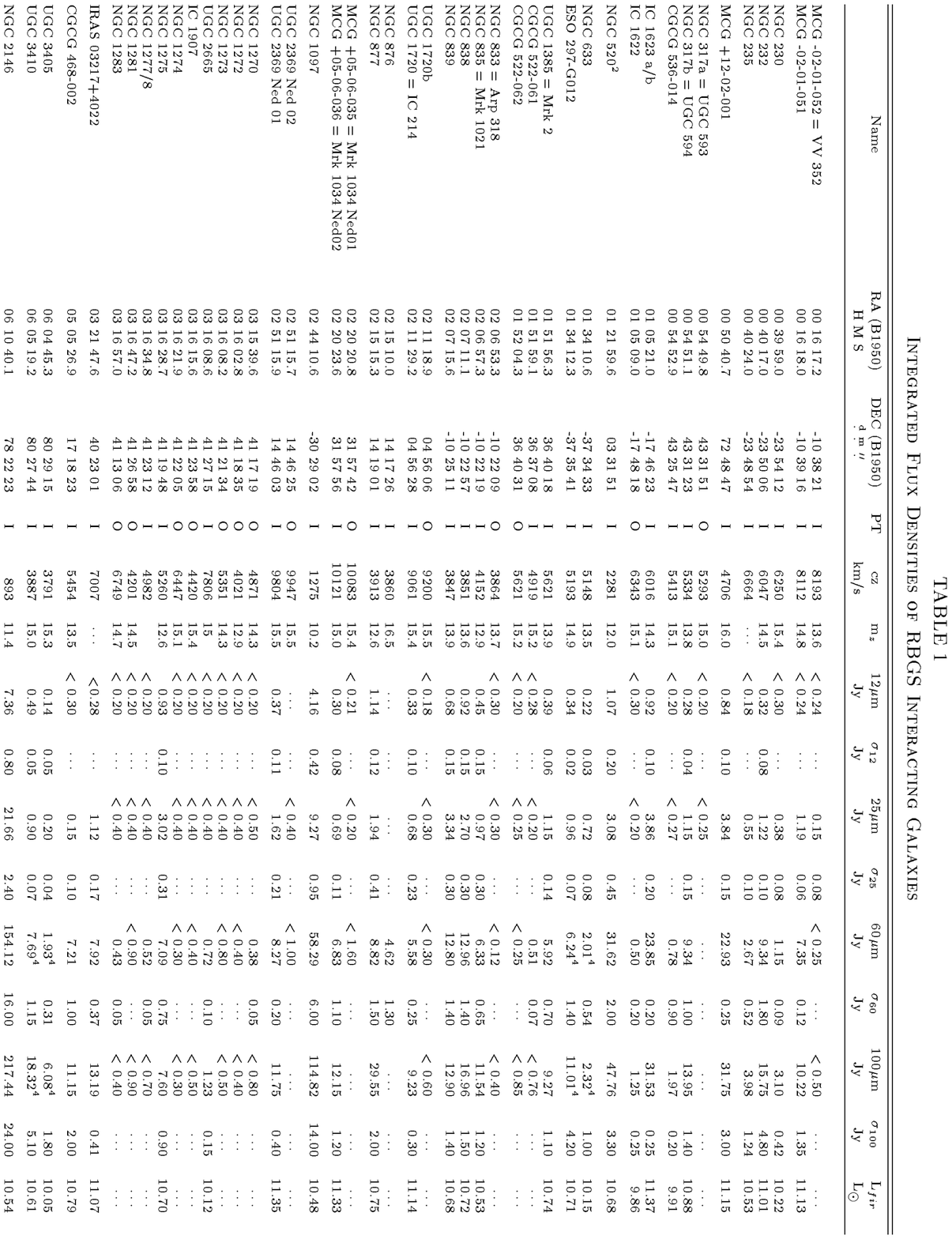} 
     \end{figure}

\clearpage  

\begin{figure}
     \plotone{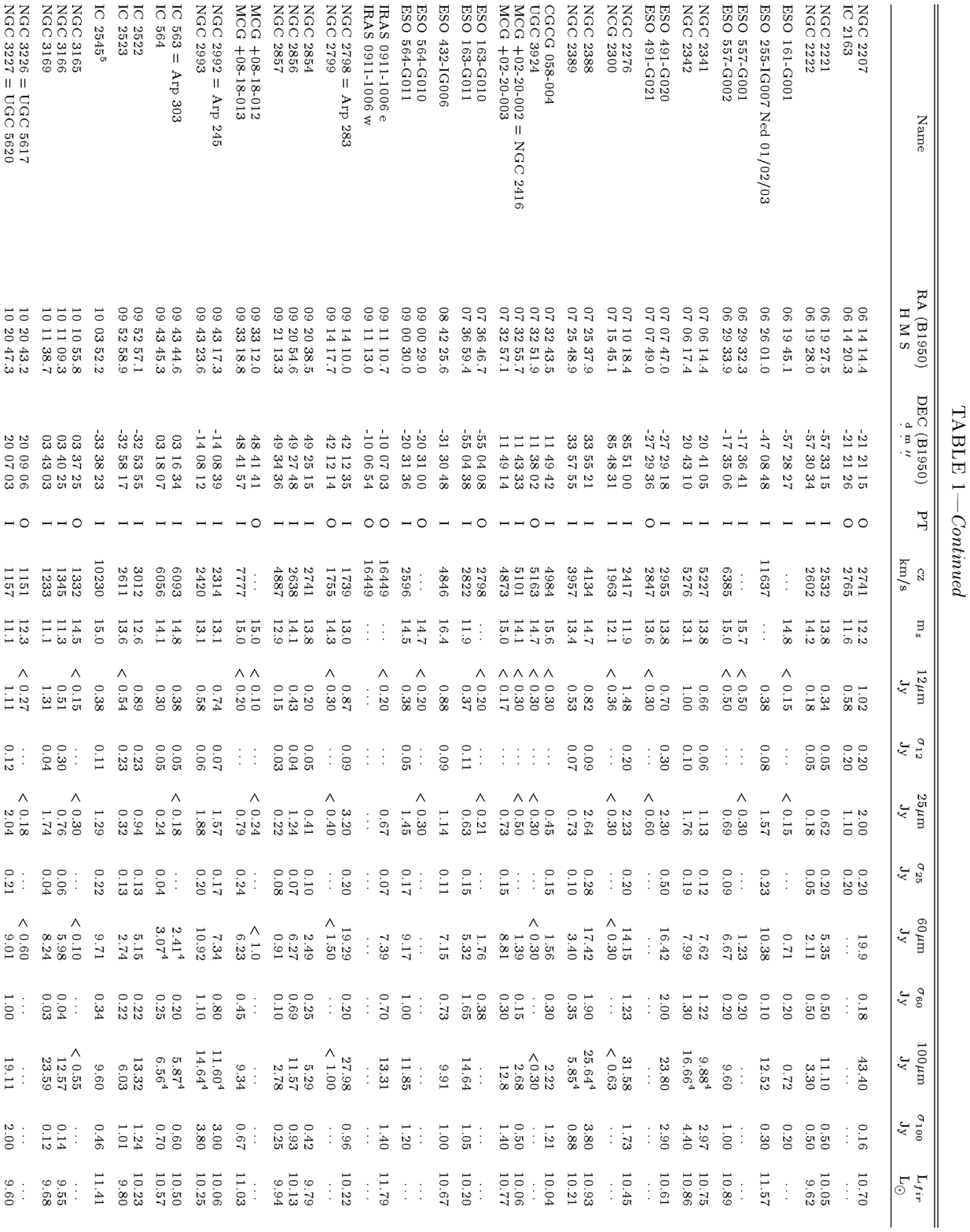} 
     \end{figure}

\clearpage  
\begin{figure}
     \plotone{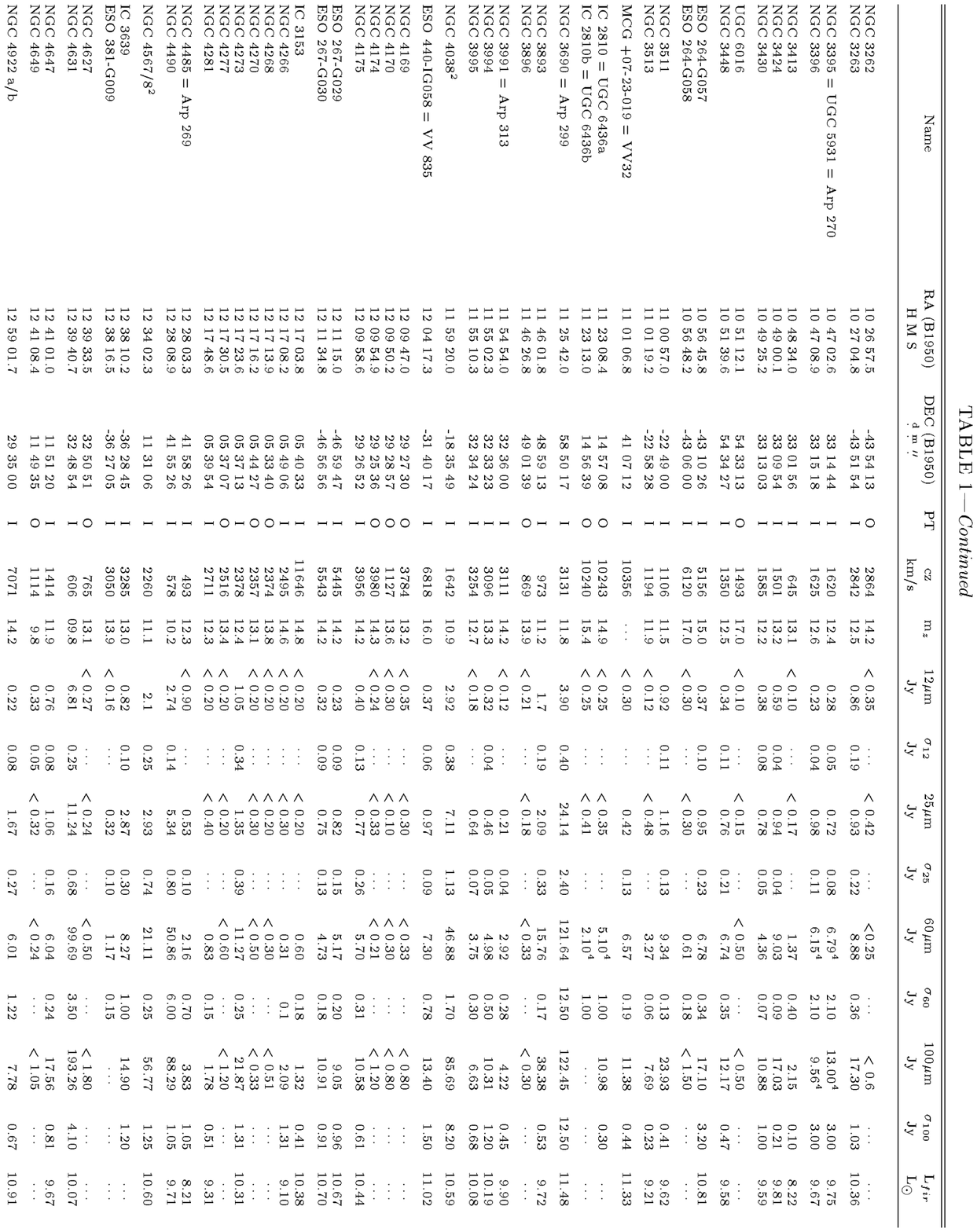} 
     \end{figure}

\clearpage  
\begin{figure}
     \plotone{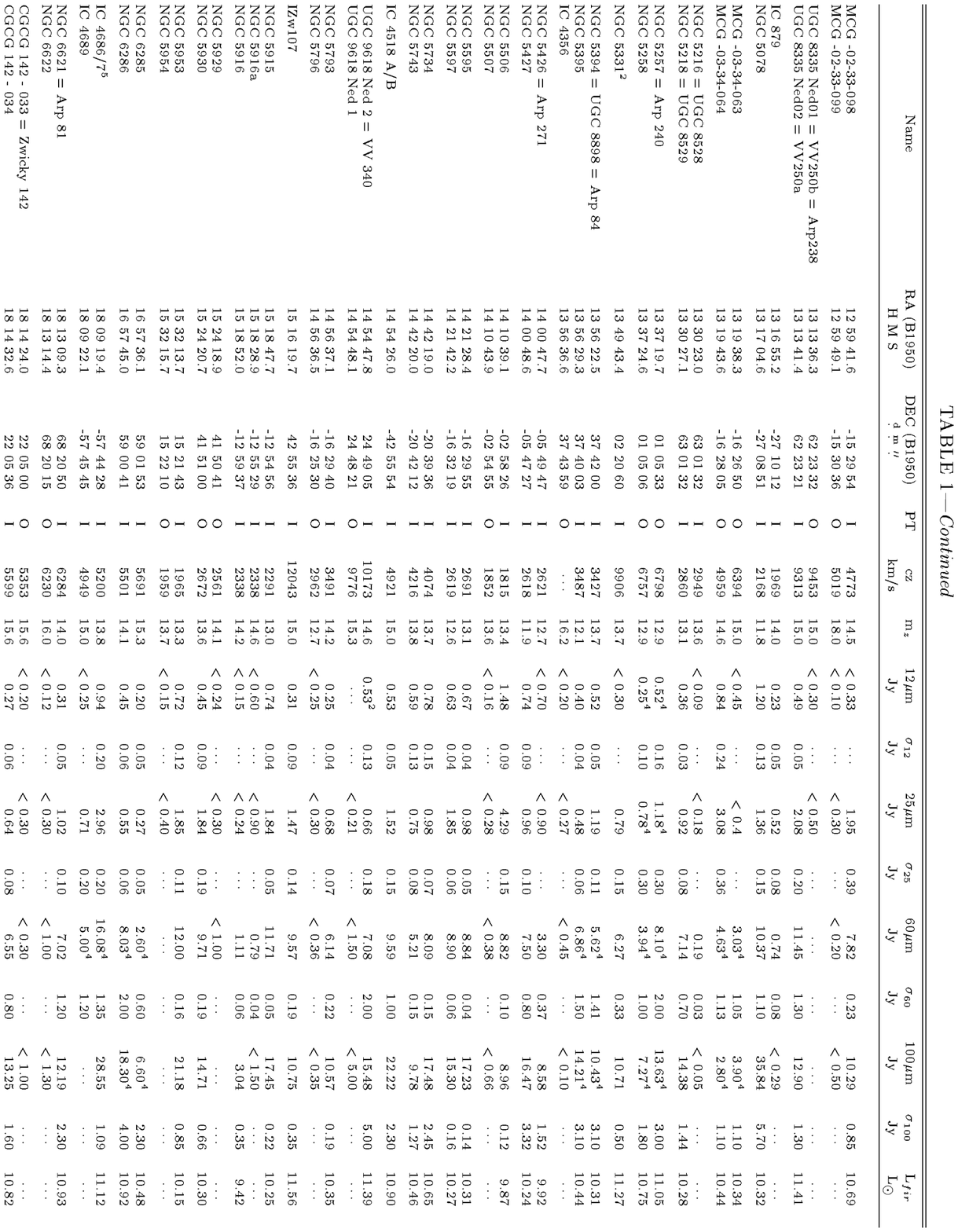} 
     \end{figure}

\clearpage  
\begin{figure}
     \plotone{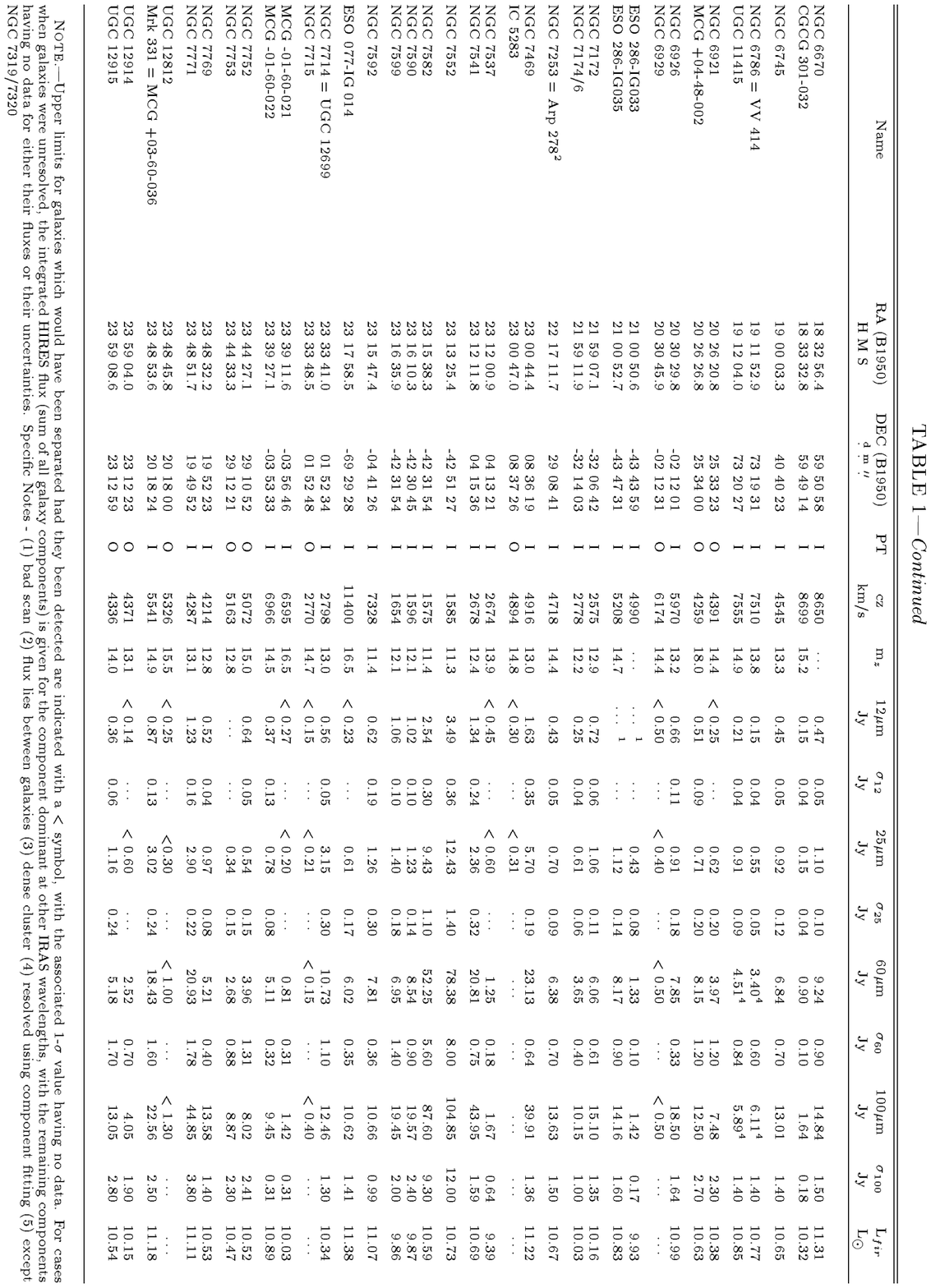} 
     \end{figure}

\clearpage  

\begin{figure}
     \plotone{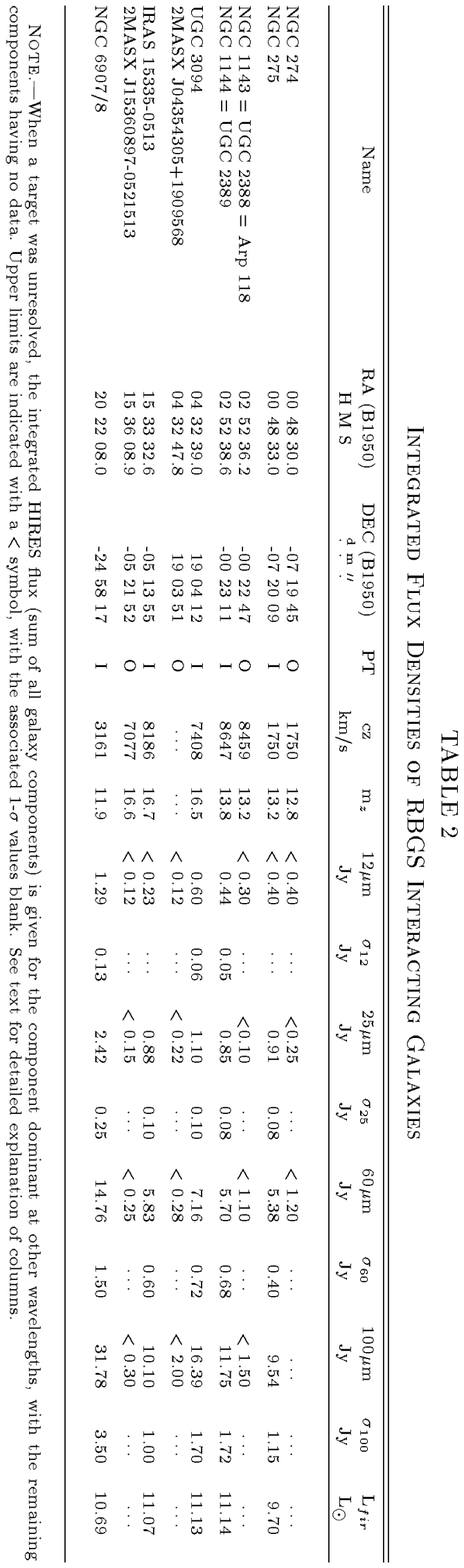} 
     \end{figure}

\clearpage  

% \begin{table}[p]
%     Due to the extreme sizes of Figure 1 and 13 and the landscaped 
%     format of Tables 1 and 2, they are not 
%     included in the text. They are available as a separate PDF 
%     download from LANL.
%     \end{table}
%    

\end{document}